\documentclass[aps, pra, 10pt,twocolumn,superscriptaddress]{revtex4-2}

\usepackage[english]{babel}
\usepackage[T2A]{fontenc}
\usepackage{lmodern}
\usepackage{graphicx}
\usepackage[caption=false]{subfig}
\usepackage{amsmath}
\usepackage{amsfonts}
\usepackage{amssymb}
\usepackage{amsthm}
\usepackage{mathtools}
\usepackage{dsfont}
\usepackage{microtype}
\usepackage{tempora}
\usepackage[usenames]{color}
\usepackage{colortbl}
\usepackage{xcolor}
\usepackage{braket}
\usepackage{hyperref}
\hypersetup{colorlinks=true, linkcolor=blue, citecolor=blue, urlcolor=blue}

\begin{document}

\newcommand{\sca}[2]{\ensuremath{\bigl({#1}\cdot{#2}\bigr)}}
\newcommand{\avr}[1]{\ensuremath{\langle{#1}\rangle}}

\newcommand{\cnj}[1]{{#1}^{\ast}}
\newcommand{\hcnj}[1]{{#1}^{\dagger}}
\newcommand{\tcnj}[1]{{#1}^{T}}
 
\newcommand{\prt}[1]{\partial_{#1}}
\newcommand{\pdrs}[1]{\partial_{#1}}
\newcommand{\pdr}[2]{\frac{\partial #1}{\partial #2}}
\newcommand{\drf}[2]{\frac{\dd #1}{\dd #2}}
\newcommand{\vdr}[2]{\dfrac{\delta #1}{\delta #2}}

 \def\BE{\begin{equation}}
 \def\EE{\end{equation}}
 \def\BA{\begin{array}}
 \def\EA{\end{array}}
 \def\BEA{\begin{eqnarray}}
 \def\EEA{\end{eqnarray}}
 \def\nn{\nonumber}
 \def\ra{\rangle}
 \def\la{\langle}
 \def\p{{\bf p}}
 \def\q{{\bf q}}
 \def\A{{\bf A}}
 \def\x{{\bf x}}
 \def\S{{\bf S}}
 \def\O{{\bf \Omega}}
 \def\I{{\bf I}}
 \def\U{{\bf U}}
 \def\V{{\bf V}}
 \def\Z{{\bf Z}}
 \def\d{\partial}

\title{Iterative $C_Z$-gate–based protocol for squeezed Schr{\"o}dinger cat state engineering}

\author{R.~Goncharov}
\email[Email address: ]{rkgoncharov@itmo.ru}
\affiliation{Quantum Information Laboratory, ITMO University, Kadetskaya Line, 3b, St. Petersburg, 199034, Russia}
\affiliation{SMARTS-Quanttelecom LLC, 6th Vasilyevskogo Ostrova Line, 59, St. Petersburg, 199178, Russia}
\author{N.~G.~Veselkova}
\email[Email address: ]{ngveselkova@outlook.com}
\affiliation{Saint-Petersburg State Institute of Technology, 26, Moskovski ave., St. Petersburg, 190013, Russia}
\author{Alexei~D.~Kiselev}
\email[Email address: ]{alexei.d.kiselev@gmail.com}
\affiliation{Quantum Information Laboratory, ITMO University, Kadetskaya Line, 3b, St. Petersburg, 199034, Russia}
\affiliation{Research and Education Center of Photonics and Optical IT, ITMO University, Kadetskaya Line, 3b, St. Petersburg, 199034, Russia} 

\begin{abstract}{Squeezed optical Schr{\"o}dinger cat states
    constitute a key resource for both fundamental tests of quantum
    theory and up-to-date quantum technologies. We propose a
    measurement-assisted gate for the generation and manipulation of
    the cat states. \textcolor{black}{In this scheme, an ancilla in the non-Gaussian small-amplitude (in general, squeezed) Schrödinger cat state and the target oscillator initially prepared in a squeezed vacuum (or coherent) state are subjected to a quantum nondemolition (QND) entangling operation followed by projective homodyne measurement.} The proposed gate enables generation of high-fidelity
    squeezed Schr{\"o}dinger cat states with controllable size and
    squeezing \textcolor{black}{with tunable fidelity/success-probability trade-off}. We also introduce an
    iterative, homodyne-conditioned $C_Z$-based protocol for cat-state
    amplification. The parameter regimes required to achieve the desired
    fidelity and the success probability are analyzed. The approach is
    well suited for applications in measurement-based quantum
    computing and hybrid quantum networks where non-Gaussian
    resources enhance computational and communication capabilities.}


\end{abstract}

\date{\today}

\maketitle

\section{Introduction}
\label{sec:introduction}

Coherent superpositions of
macroscopically (classically) distinguishable quantum
states
are known as the
Schr{\"o}dinger cat states (SCSs)~\cite{Leonhardt_1995, Leggett_2002,
  Lvovsky_2013}.
In quantum optics, two-component photonic SCSs are typically realized
as superpositions of
the coherent states of the electromagnetic field
that, without the loss of generality,
can be written in the following form~\cite{Sanders2012, Lvovsky2020}
\begin{align}
\label{cat1}
|{\rm cat}_{\varphi}(\alpha)\rangle = \frac{1}{\sqrt{2N_{\varphi,\alpha}}} \big(|\alpha\rangle + e^{i\varphi}|-\alpha\rangle \big),
\end{align}
where
$\alpha$ is the displacement amplitude;
$\varphi$ is the
relative phase between the two components
and
$N_{\varphi,\alpha} = 1 + e^{-2|\alpha|^2}\cos\varphi$ is the
normalization factor.
\textcolor{black}{Macroscopic distinguishability requires the separation of the two
coherent components to exceed the vacuum (shot-noise) width. In the
quadrature convention used below, this separation is $2a$ with
$a=\sqrt{2}|\alpha|$; throughout the paper, $a$ is used as the
cat-state size parameter.}

Such states are known to
play a central role in both fundamental tests of quantum
theory~\cite{Brune_1996, Wenger_2003, Garc_a_Patr_n_2004} and as an
important resource for upcoming quantum information technologies.
The latter include quantum computation~\cite{Guillaud_2023, Ralph_2003,
  Gilchrist_2004, Lund_2008},
metrology~\cite{Joo_2011,Facon2016,Knott2016,Duivenvoorden2017},
quantum teleportation
protocols~\cite{Enk2001,Lee2013},
error
correction schemes~\cite{Weigand2018, Hastrup2020, Cai2021,
  Chamberland2022}, quantum communication and
repeaters~\cite{vanLoock2008,Brask_2010, Sangouard_2010,
  Goncharov2022}.

Squeezed SCSs represented by linear superpositions of displaced squeezed coherent
states
can be regarded as a generalization the coherent-state SCSs~\eqref{cat1}
and exhibit enhanced quantum properties~\cite{Shankar2011,
  Kim2005} such as improved phase sensitivity and robustness against
dephasing and photon loss. These features make them particularly
attractive for bosonic encoding and fault-tolerant quantum information
processing~\cite{DaSilva2016, Wei2012}.

Though several ideas have been proposed to create squeezed optical
SCSs of propagating
light~\cite{Ourjoumtsev_2007, Etesse_2015, Huang_2015},
efficient generation of high-fidelity
large-photon-number SCSs remains a challenging problem in quantum
optics. Since the optical SCSs are non-Gaussian bosonic states,
their generation demands that
the input states are either non-Gaussian~\cite{Sychev_2017, Ourjoumtsev_2007, Lund_2004, Laghaout_2013}
or they undergo a non-Gaussian
evolution~\cite{Weedbrook_2012, Chabaud_2020}.

Strong optical nonlinearity for squeezed cat-state generation can be
induced by a non-Gaussian measurement such as photon-number
detection~\cite{Dakna_1997,Ourjoumtsev_2007, Takahashi_2008,
  Takase_2021, Eaton_2022}, or can be based on the evolution of the
initial state under a strong nonlinear interaction such as the
spontaneous parametric down-conversion process~\cite{Ourjoumtsev_2007,
  marek2010coherent, ourjoumtsev2009generation, lvovsky2018squeezed,
  liu2020generation, mazzarella2021heralded, wang2022generation,
  zhang2023generation, chen2022progress} or the Kerr
effect~\cite{Yurke_1986, Olsen2005, Takeda2019}. Enhanced non-Gaussian
features are achieved by engineered interactions like strong
nonlinearities or specific gate operations that produce non-Gaussian
states regardless of measurement outcomes. Alternatively, in some
schemes, the measurement process is integrated into the operation so
that the overall process is effectively
deterministic~\cite{Takagi2018}.

In alternative approaches, the non-Gaussian features are typically
generated via the evolution of the system, especially through
measurement-based post-selection or conditional operations,
non-Gaussian ancillas, entangling gates, and weak nonlinear
interactions. Many conditional schemes rely on well-established
techniques such as photon subtraction~\cite{Jeong_2002,
  ourjoumtsev2009generation, Ourjoumtsev2016, Yoshikawa2018,
  Takahashi2010} and homodyne detection, making them more feasible
with current optical technologies compared to deterministic methods
that often require challenging
nonlinearities~\cite{Ourjoumtsev2013}. Note that conditional schemes
for generating squeezed SCSs using non-Gaussian input states offer
advantages~\cite{Jeong2009, Ourjoumtsev_2006, Takagi2018,
  Neergaard_Nielsen_2010, Ourjoumtsev2013} over deterministic schemes
based on strong nonlinear interaction, such as higher state fidelity,
reduced experimental complexity by avoiding strong nonlinear
interactions, greater flexibility in state engineering, and improved
feasibility with current optical technologies.

Most quantum information applications require SCSs made of coherent states
with reasonably high displacement amplitudes~\cite{Gilchrist_2004,
  Walmsley_2015}. Typically, a moderate amplitude modulus
$|\alpha|$, corresponding to a mean photon number of a few photons
($\langle n\rangle=|\alpha|^2$), might suffice for applications like
quantum key distribution where the entanglement properties are
more important than the displacement amplitude, whereas
larger-amplitude cat states (containing 10 or more photons) are
preferred for quantum error-correcting codes~\cite{Gottesman2001,
  Vasconcelos2010, Weigand2018, Hastrup2020}, and fault-tolerant
quantum computing~\cite{Ralph_2003, Cochrane_1999}.
Large-amplitude cat states can provide better performance and
precision measurements, but are more challenging to generate and more
susceptible to decoherence. The optimal displacement amplitude depends on the
balance between the complexity of the generation and the specific
requirements of the intended applications.


Recent years have seen significant progress in protocols for iterative
amplification of SCSs from basic probabilistic schemes to
sophisticated, adaptive, and machine learning-assisted iterative
protocols~\cite{marek2018generation, Wang2020, Miranowicz2020} to
produce larger superpositions with improved fidelity~\cite{Biagi_2022,
  Winnel_2024, Eaton_2022, Hutin_2025}. Such protocols taking
advantage of novel experimental techniques and theoretical frameworks
include iterative multi-step cat breeding~\cite{Zhang_2022,
  Solodovnikova2025}, iterative amplification of SCSs in
superconducting circuits~\cite{Vlastakis2019, Zhang_2022,
  zhang2023generation}, amplification protocol based on a frequency
comb~\cite{Song_2022}, noiseless linear amplification~\cite{Zhang2018,
  Gao2023}, photon addition and subtraction techniques in cavity QED
systems~\cite{liu2020generation}, on-chip nonlinear amplification
methods, scalable quantum network approaches.


In early conditional experiments based on single-photon subtraction
from squeezed light, strongly non-Gaussian ``kittens'' with Wigner
negativity were reconstructed tomographically, but typically
benchmarked against unsqueezed SCS targets and with modest amplitudes
($|\alpha|\approx 1$)~\cite{Ourjoumtsev_2006,NeergaardNielsen_2006,Wakui_2007,Dong_2014}. Access
to genuinely squeezed SCSs required higher-order non-Gaussian
resources. A landmark demonstration used a heralded two-photon Fock
state together with linear optics and homodyne conditioning to
generate an even squeezed cat~\cite{Ourjoumtsev_2007}.
Full tomography revealed the expected
phase-space interference of two coherent components with a target
model corresponding to the squeezed SCS of mesoscopic size and
approximately $3$~dB of squeezing.

Subsequent multi-photon-assisted and breeding protocols increased the
achievable size. Interfering with two single-photon states and
conditioning on a homodyne outcome produced an even squeezed cat with
$|\alpha|=1.63$ and fidelity $F=0.61$ to the ideal
target~\cite{Etesse_2015}. Using photon-number-resolving transition
edge sensor detectors to subtract multiple photons from a squeezed
vacuum pulse, three-photon subtraction yielded a SCS with
$\langle n\rangle=2.75$ and fidelity $F=0.59$ to an ideal odd SCS,
illustrating the scaling of the amplitude scale with heralded photon
number~\cite{Gerrits_2010}. A complementary heralded route based on
two-mode squeezed vacuum and two-photon detection reported propagation
of even squeezed cats with $|\alpha|^2\approx3$ and fidelity
$F\approx0.67$ to the ideal squeezed SCSs~\cite{Huang_2015}. In a cat
breeding implementation, an even squeezed cat of $|\alpha|\approx1.85$
and squeezing $\sim 3$ dB was obtained with $F\simeq0.77$, and with a
success probability of $\sim0.2$~\cite{Sychev_2017}.

Deterministic manipulation of traveling cat states has also advanced
significantly. An in-line all-optical squeezer applied to previously
prepared cat states produced squeezed-cat outputs with the amplitudes
$|\alpha|\simeq0.99$–$1.40$, the squeezing parameters $r\simeq0.24$–$0.30$
(2.1–2.6 dB) and the fidelities $F\simeq0.61$–$0.65$, at kilohertz rates
determined by the input source~\cite{Wang_2022}. Winnel \textit{et
  al.}~\cite{Winnel_2024} theoretically analyzed two fully
deterministic schemes employing Gaussian optics with Fock-state
ancillas. Using 6~dB of inline or ancillary squeezing and detectors
with efficiency $\epsilon\simeq0.98$, their simulations for $n=10$
Fock inputs produced large-amplitude squeezed cats with component
separation $|\alpha|\simeq3$ and theoretical fidelities exceeding 0.9
for ideal detectors and around 0.6 for nonideal ones. The anticipated
squeezing for such parameters is approximately 6~dB, and increasing
the Fock number further drives $F\to1$, demonstrating fully
deterministic GKP-compatible squeezed-cat generation. Zhang \textit{et
  al.}~\cite{Zhang_2022} proposed an all-optical three-mode scheme in
which an engineered two-photon pump and a Kerr-type interaction
autonomously stabilize a single-mode squeezed SCS. For representative
parameters $|\alpha|=2$ and more than 9 dB of theoretical squeezing,
the steady state achieves $F\simeq0.958$ and approaches unit fidelity
in the lossless limit, establishing an experimentally feasible route
to deterministic optical SCS generation.

Beyond free-space optics, stabilized cat manifolds and protected
rotations have been demonstrated in circuit-QED
platforms~\cite{Pan_2023}, underscoring the broader utility of
squeezed SCSs for fault-tolerant control.

The above-discussed studies taken together delineate the presently
achievable optical squeezed-cat parameters for propagating
fields. These figures define realistic targets for new
measurement-assisted gate schemes aimed at producing undistorted
squeezed SCSs of controllable size and squeezing, with higher success
probabilities where applicable, which constitutes the main focus of
the present work.

In this paper, we propose a scheme for the conditional generation of
large-amplitude squeezed SCSs with a controllable degree of squeezing,
starting from an arbitrary coherent state and achieving a high
probability of success.
The approach employs a two-node non-Gaussian
gate where a small-amplitude Schr{\"o}dinger cat ancilla and
the target oscillator prepared in
the vacuum (or arbitrary coherent) initial state 
are subjected to
a quantum non-demolition (QND) entangling operation ${\hat C}_z$
followed by a projective homodyne measurement. The gate parameters
that enable the generation of squeezed SCSs with the desired fidelity
and success probability are analytically and numerically evaluated.

Building upon the considered gate, we develop a homodyne-conditioned
iterative cat-state amplification protocol that enables controlled
growth of both the coherent-state separation and the squeezing of the
output state at each iteration. This approach facilitates the
preparation of squeezed Schr{\"o}dinger cat–like states with desired
properties, offering a flexible resource for continuous-variable
quantum information processing.

\section{State at gate output}
\label{section_II}

\begin{figure}[t!]
\centering
\includegraphics[width=1.0\columnwidth]{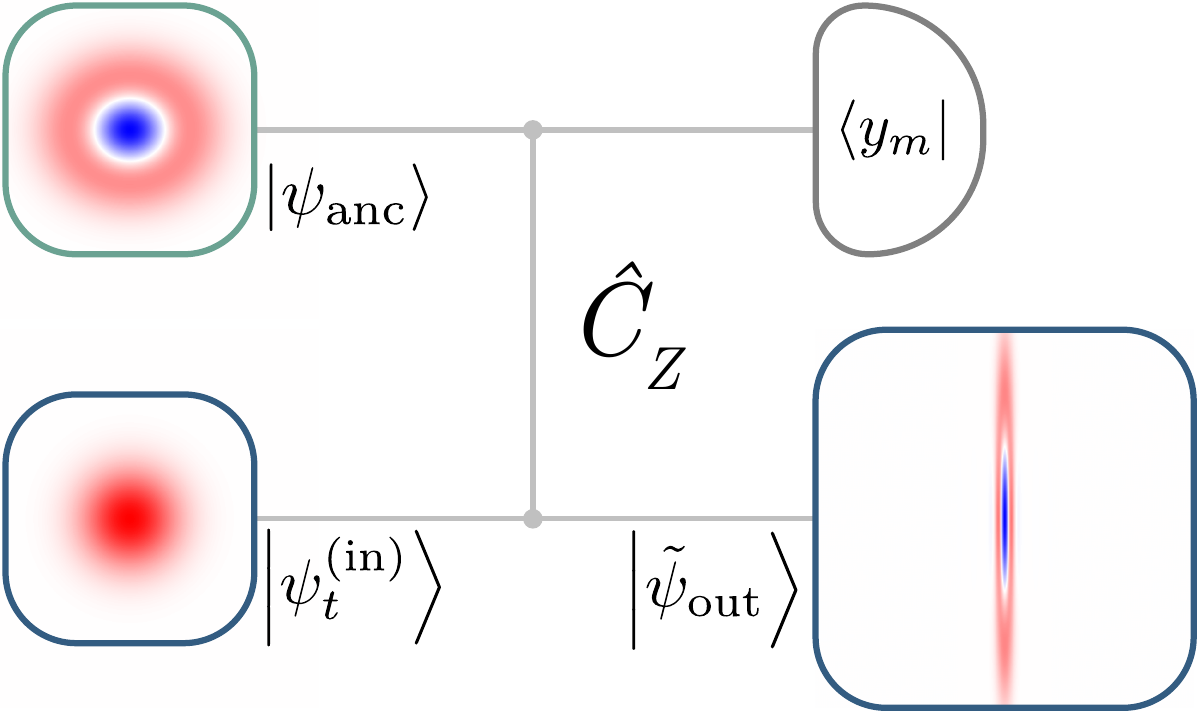}
\caption{{The scheme for conditional generation of squeezed optical
    SCSs based on the small-amplitude SCS of the ancilla
    $|\psi_{\rm anc}\rangle$ as an elementary non-Gaussian resource, a
    vacuum state (or an arbitrary coherent state) of the target
    oscillator $|\psi^{\rm (in)}_t\rangle$ and a two-mode quantum
    nondemolition entangling operator
    $\hat{C}_Z= \exp{(iG\hat{q}_1\hat{q}_2)}$, followed by the
    projective homodyne measurement $\bra{y_m}$ of the ancilla
    momentum.}}
\label{figL1}
\end{figure}

{Here we consider the measurement-assisted two-node logic gate shown
  in Fig.~\ref{figL1} which is based on the continuous-variable (CV)
  controlled-Z operation $\hat{C}_Z$~\cite{Wang_2011, Alexander_2017,
    Sakaguchi_2023, Matsos_2025}, also called a quantum nondemolition
  (QND) gate, which is a central entangling operation in various
  applications of CV quantum information processing, and projective
  homodyne measurement, for the resource Schr{\"o}dinger cat state
  (SCS) of the ancillary oscillator.}

{We assume the target oscillator is initially prepared in a squeezed
  coherent state with complex amplitude
  $\alpha = x_0/\sqrt{2} +ip_0/\sqrt{2}$,
\begin{align}
\label{a0}
|\psi^{\rm (in)}_t\rangle = \int dx_1\,\psi_{in}(x_1)|x_1\rangle,
\end{align}
where 
\begin{align}
\label{a1}
&
    \psi_{in}(x_1)\equiv\psi_{\delta,x_0,p_0}(x_1)   
  \notag
  \\
  &
 = \frac{\sqrt{\delta}}{\pi^{1/4}} \exp\Big\{-\frac{\delta^2(x_1-x_0)^2}{2}+ip_0x_1-\frac{ip_0x_0}{2}\Big\},
\end{align}
is the wave function in the position-representation;
$x_1$ is the coordinate of the target
oscillator; and
$\delta^{-1}$
is the x-quadrature squeezing factor~\cite{Walls1983} defined as the
ratio of the x-quadrature standard deviation in the squeezed state to
the deviation in the coherent state (the corresponding variance ratio is
$\delta^{-2}$).
In the context of squeezing expressed in dB units, the
quantity usually quoted is the x-quadrature variance squeezing level:
\begin{align}
  \label{a01}
  &
\text{X-quadrature squeezing level (in dB)}
  = 10\log_{10}(\delta^{-2})
  \notag
  \\
  &
  = -20\log_{10}(\delta),
\end{align}
thus, a negative squeezing level (in dB) corresponds to the squeezing
in the x-quadrature, and a positive one to the anti-squeezing in it.}

{An ancillary oscillator is prepared in a squeezed even/odd SCS aligned along the x-axis in phase space, 
\begin{align}
\label{a2}
|\psi_{\rm anc}\rangle = \int dx_{2}\,\Psi^{\pm}_{r,a,0}(x_2)|x_{2}\rangle,
\end{align}
so that the ancilla momentum measurement outcome could be compatible
with two clearly distinguishable values of the ancillary coordinate
\footnote{Within the semiclassical description, where a bipartite SCS
  is represented by two points in phase space, one can conclude that
  for a horizontally oriented SCS (i.e., extended along the x-axis),
  the measurement outcome of the ancillary oscillator's momentum may
  correspond to two distinct values of its coordinate. By means of the
  entangling operation $C_Z$, which couples the target and ancillary
  subsystems, the coordinate ambiguity present prior to measurement is
  transferred to the momentum of the target oscillator. As a
  consequence, a state of quantum superposition (a cat-like state)
  emerges in which the interfering components are mutually displaced
  along the momentum variable. Further details of the semiclassical
  geometric approach are provided
  in~\cite{Sokolov_2020,Veselkova2026}}. Here, the
position-representation wave function has the form (in terms of the
function of Eq. (\ref{a1}))
\begin{align}
\label{a3}
\Psi^{\pm}_{r,a,0}(x_2)= \mathcal{N}_{\pm}[\psi_{r,a,0}(x_2)\pm \psi_{r,-a,0}(x_2)],
\end{align}
with the normalization factor
\begin{align}
\label{a4}
\mathcal{N}_{\pm}=[2(1\pm\exp{(-r^2 a^2)})]^{-1/2},
\end{align}
where $a$ is the cat-state size parameter in the $x$ quadrature; for an
unsqueezed coherent-state component it is related to the coherent
amplitude by $a=\sqrt{2}|\alpha|$. In what follows, ``size'' refers to
this amplitude/separation parameter or to its transformed value (for
example, $|G|a$ or $|G|^k a$), and not to the mean photon number.
}

{A two-mode entangling QND operator $\hat{C}_Z= \exp{(iG\hat{q}_1\hat{q}_2)}$ characterized by the dimensionless coupling constant $G$ between two bosonic modes is applied to the initial state $|\psi_t^{(\mathrm{in})}\rangle\otimes|\psi_{\rm anc}\rangle$ of the composite system, 
resulting in the state given by}
{
\begin{align}
  \label{a5}
  &
    \hat{C}_Z\ket{\psi_t^{\rm(in)}}\otimes\ket{\psi_{\rm anc}}=\int \psi_{in}(x_1)\Psi^{\pm}_{r,a,0}(x_2)  
  \notag
  \\
  &
  \times e^{i Gx_1 x_{2}}\left|x_1\right\rangle\otimes\left|x_{2}\right\rangle\, d x_1 d x_{2}.
\end{align}
}

{A subsequent projective ancilla momentum measurement performed on the state~\eqref{a5} with the outcome $p^{\rm (out)}_2=y_m$ and the corresponding momentum eigenstate}
{
\begin{align}
  \label{a6}
  &
    \ket{y_m}=\frac{1}{\sqrt{2\pi}}\int e^{i y_m x_2}\ket{x_2} d x_2
\end{align}
}{
results in a reduction of the total quantum state to the next output state:}
{\begin{align}
  \label{a7}
  &
    \ket{\tilde{\psi}_{\rm out}}=\bra{y_m}\hat{C}_Z\ket{\psi_t^{\rm(in)}}\otimes\ket{\psi_{\rm anc}}
    \notag
  \\
  &
    = \int\psi_{in}(x)\,\widetilde{\Psi}^{\pm}_{r,a,0}(y_m-Gx)\ket{x} dx,
\end{align}}
{where we have used the relation
$\avr{y_m|x_2}=e^{-iy_m x_2}/\sqrt{2\pi}$ and the standard momentum-representation wave function $\widetilde{\Psi}^{\pm}_{r,a,0}$
given by}
{
\begin{align}
  &
    \widetilde{\Psi}^{\pm}_{r,a,0}(p)=\frac{1}{\sqrt{2\pi}}
    \int e^{-i p x_{2}} \Psi^{\pm}_{r,a,0}(x_{2})\, d x_2
    \notag
  \\
  &
   \label{a10}
  =\mathcal{N}_{\pm}\,\frac{1}{\pi^{1/4}\sqrt{r}}\,e^{-p^2/(2r^2)}(e^{-ipa}\pm e^{ipa}).
 \end{align}}
{Equation~(\ref{a7}) shows that the output wave function in the position representation is the product
of the input wave function and a gate-added factor -- the momentum-space wave
function of the resource state evaluated at momentum $y_m-Gx$:
\begin{align}
\label{a011}
\tilde{\psi}_{out}(x,y_m)= \psi_{in}(x)\,\widetilde{\Psi}^{\pm}_{r,a,0}(y_m-Gx).
\end{align}}
{The explicit expression of this (non-normalized) wave function has the form, up to an unimportant phase factor,}
{\begin{align}
 &
    \label{a012}
    {\tilde{\psi}}_{\rm out}(x,y_m) = \mathcal{N}_{\pm}\,\sqrt{\frac{\delta}{\pi r}}\,e^{-\delta^2(x-x_0)^2/2}\,e^{-(Gx-y_m)^2/(2r^2)}
\notag
  \\
  &
  \times
\Big(e^{ix(p_0+Ga)}\pm e^{i2y_m a}e^{ix(p_0-Ga)}\Big).    
\end{align}}

{From Eq. (\ref{a012}), it follows that at the gate output, we obtain the heralded ``perfect'' (undistorted) squeezed SCS whose components are spaced in momentum by $2|G|a$, provided that $y_m=Gx_0$. In this case, the output squeezed-cat-state wave function is}
{\begin{align}
  &
\psi^{\rm cat}_{\rm out}(x) =\frac{1}{\sqrt{\mathcal{N}}}{\tilde{\psi}}_{\rm out}(x),
    \label{a11}
  \\
  &
    \label{a12}
    {\tilde{\psi}}^{\rm cat}_{\rm out}(x) = \mathcal{N}_{\pm}\,\sqrt{\frac{\delta}{\pi r}}\exp\!{\Big[\!-\!\frac{(\delta^2+(G/r)^2)(x-x_0)^2}{2} \Big]}
\notag
  \\
  &
  \times
\Big(e^{ix(p_0+Ga)}\pm e^{i\varphi_{\rm out}}e^{ix(p_0-Ga)}\Big),
\end{align}
with the relative phase between cat components $\varphi_{\rm out}\equiv 2Gx_0a$; here ${\tilde{\psi}}_{\rm out}(x)$ is the unnormalized output wave function of the target oscillator and $\mathcal{N}$ is the normalization factor. As can be seen, the quadrature squeezing factor of the output SCS is $\gamma^{-1}$, where the coefficient
\begin{align}
\label{a13}
\gamma\equiv\sqrt{\delta^2+(G/r)^2}
\end{align}
is expressed in terms of the QND coupling coefficient $G$ and the squeezing factors of the target $\delta^{-1}$ and ancillary $r^{-1}$ states, and the components of the SCS are shifted along the $p$ axis by $\pm Ga$.}

{For centered states ($x_0=p_0=0$), Eq.~(\ref{a12}) shows that the cat state at the gate output has the same parity as the ancillary input cat when $\varphi_{\rm out}=2\pi k$ and the opposite parity when $\varphi_{\rm out}=\pi(2k+1)$, $k\in Z$. For displaced states, these conditions determine the relative phase and sign of the two output components rather than a strict parity eigenvalue about the origin.}

\begin{figure*}
\centering
  \begin{tabular}{c @{\qquad} c @{\qquad} c}
    \includegraphics[width=0.5\linewidth]{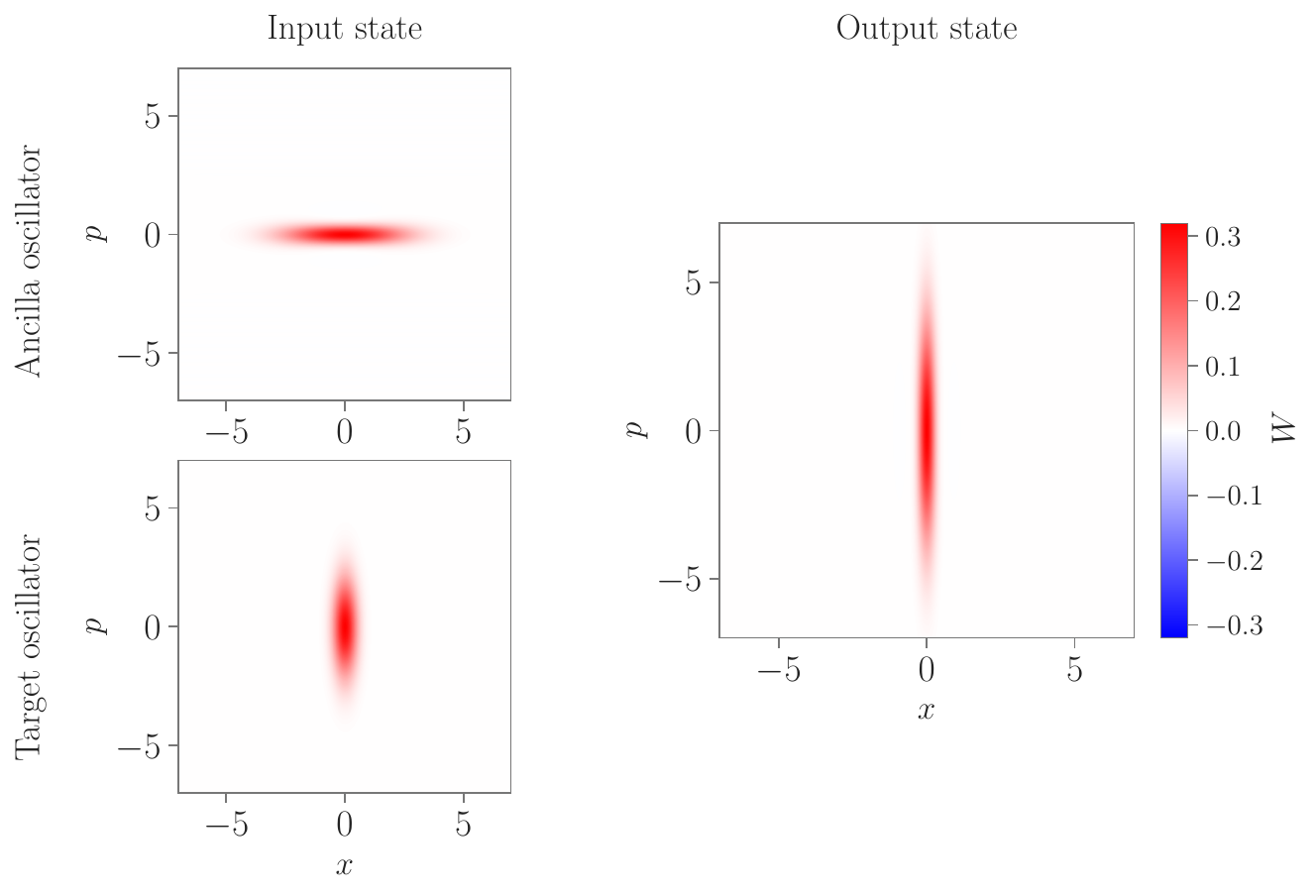} &
    \includegraphics[width=0.5\linewidth]{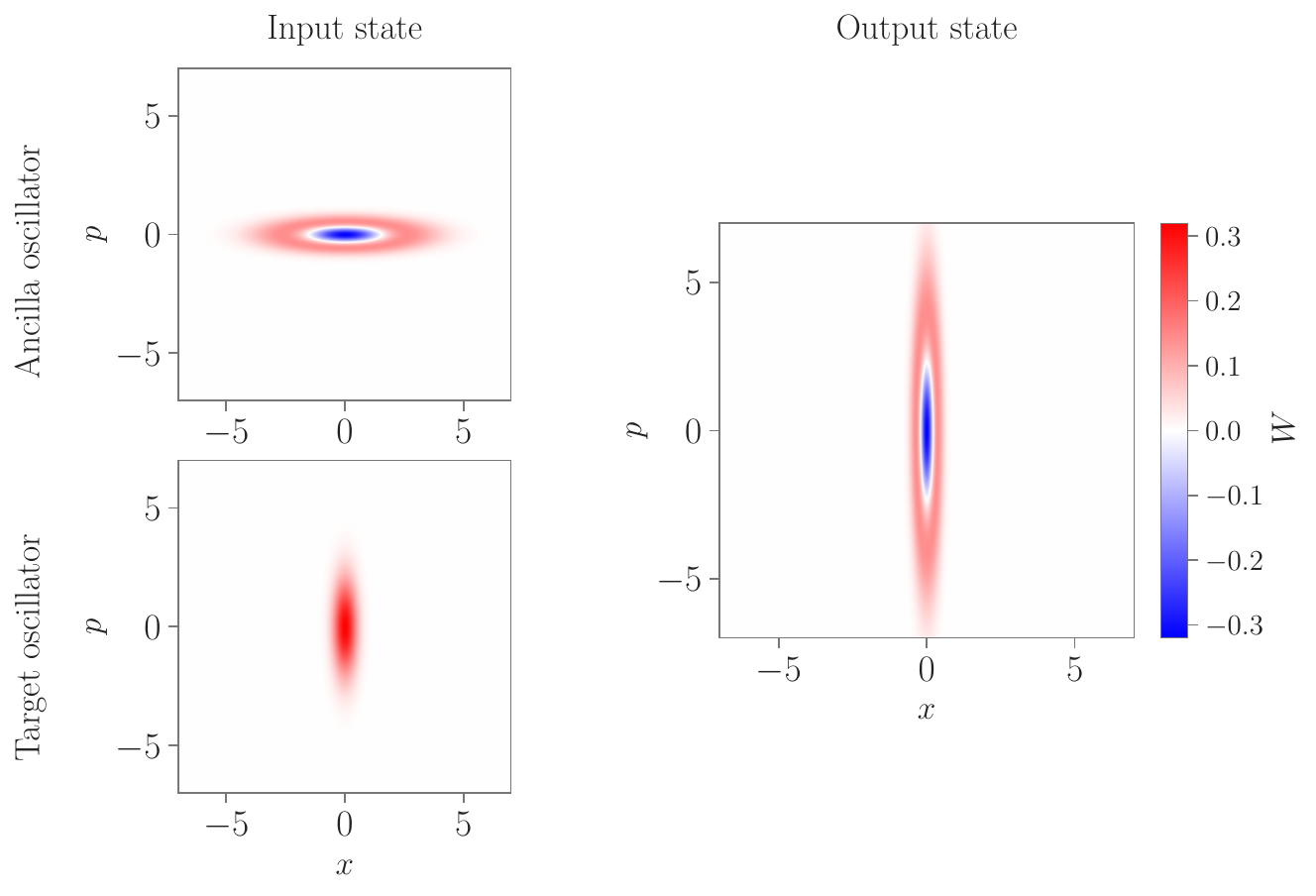}\\
    \small (a) & \small (b)
  \end{tabular}
  \caption{{Wigner functions of the quantum states at the input and output of the investigated gate for parameter values $\delta=2.0$, $r=0.5$, $G=1.2$, $a=\sqrt{2}$. The target oscillator is prepared in a vacuum state squeezed along the x-quadrature ($\delta=2.0$), while the ancillary oscillator is in either (a) an even or (b) odd SCS, both with the amplitude $\alpha=a/\sqrt{2}=1$ and squeezed along the p-quadrature ($r=0.5$). The resulting output state is an (a) even or (b) odd Schr{\"o}dinger cat state squeezed in the x-quadrature ($\gamma\approx3.12$) with an effective size $|G|a\approx 1.70$.  The color scale
    indicates the Wigner function ranging from $-0.3$ (blue) to $0.3$
    (red).}} 
\label{figB1}
\end{figure*}

\begin{figure*}
\centering
  \begin{tabular}{c @{\qquad} c @{\qquad} c}
    \includegraphics[width=0.5\linewidth]{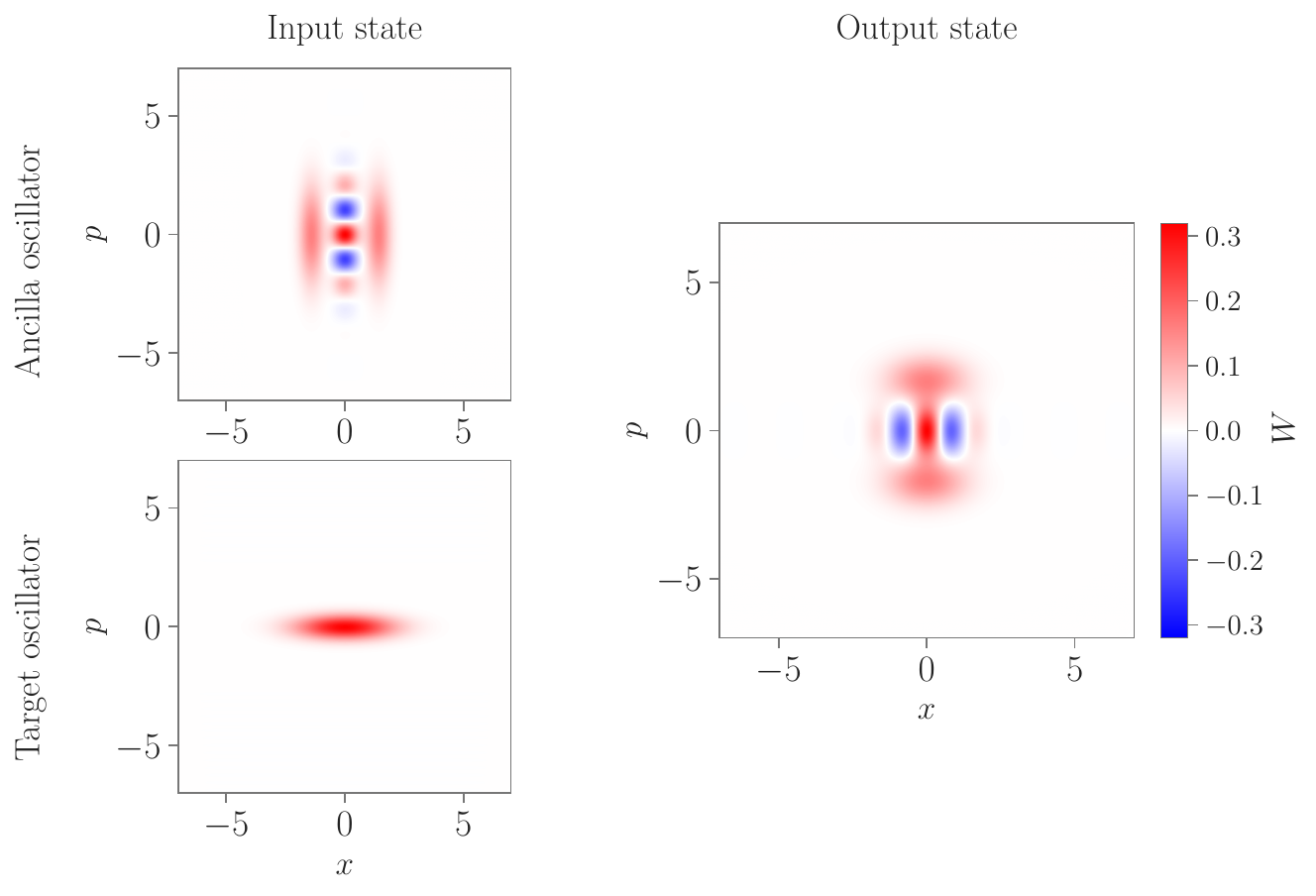} &
    \includegraphics[width=0.5\linewidth]{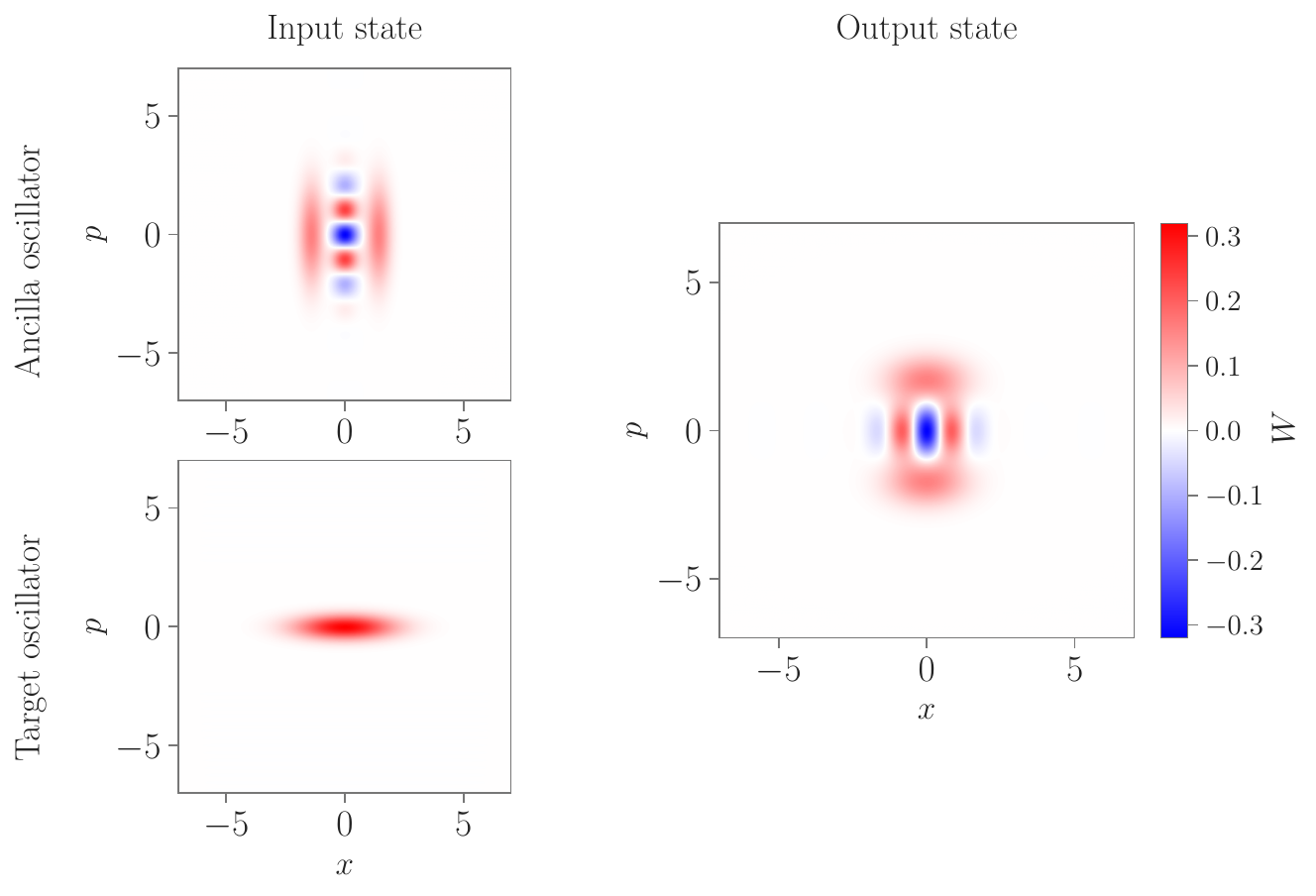}\\
    \small (a) & \small (b)
  \end{tabular}
  \caption{{Wigner functions of the quantum states at the input and output of the investigated gate for parameter values $\delta=0.5$, $r=2.0$, $G=1.2$, $a=\sqrt{2}$. The target oscillator is prepared in a vacuum state squeezed along the p-quadrature ($\delta=0.5$), while the ancillary oscillator is in either (a) an even or (b) odd SCS, both with the amplitude $\alpha=a/\sqrt{2}=1$ and squeezed along the x-quadrature ($r=2.0$). The resulting output state is an (a) even or (b) odd Schr{\"o}dinger cat state squeezed in the p-quadrature ($\gamma\approx 0.78$) with an effective size $|G|a\approx 1.70$. The color scale
    indicates the Wigner function ranging from $-0.3$ (blue) to $0.3$.}} 
\label{figB2}
\end{figure*}

{As follows from the above expressions, the proposed gate can generate SCSs exhibiting tunable squeezing in either the x-quadrature (for $\gamma > 1$) or the p-quadrature (for $\gamma < 1$), controlled by the parameter $\gamma$. Both cases are illustrated in Fig.~\ref{figB1} (x-quadrature squeezing of the output state) and Fig.~\ref{figB2} (p-quadrature squeezing) through the construction of the input and the output Wigner functions%
\footnote{\begingroup\color{black}For a pure state with coordinate wave function $\psi(x)$ we plot
the Wigner function
\begin{align}
\label{p0}
W(x,p)=\frac{1}{\pi}\!\int dz\,\psi^{*}(x+z)\psi(x-z)e^{2ipz},
\end{align}
with $\hbar=1$. Semiclassically, the two branches of a horizontally
oriented ancillary SCS are centered near
$(x_A,p_A)=(\pm a,0)$. The QND gate obeys
$\hat C_Z^\dagger \hat p_T\hat C_Z=\hat p_T+G\hat x_A$ and
$\hat C_Z^\dagger \hat p_A\hat C_Z=\hat p_A+G\hat x_T$; hence an
ancilla momentum outcome $y_m\simeq Gx_0$ leaves the two ancilla
branches indistinguishable while mapping them to target momenta
$p_0\pm Ga$. The output is therefore a squeezed cat whose components are
separated by $2|G|a$ in the target momentum quadrature; equivalently,
its output size parameter is $|G|a$. Outcomes
$y_m\ne Gx_0$ introduce the Gaussian distortion and relative phase
shown explicitly in Eq.~(\ref{a012})\endgroup}}.


{The statistics of the homodyne measurements of the ancilla momentum is described by the probability density to observe the outcome $y_m$ given by the norm of the unnormalized output wave function~(\ref{a012})  
\begin{align}
  &
\label{a015}
P(y_m)=\langle{\tilde{\psi}}_{\rm out}|{\tilde{\psi}}_{\rm out}\rangle=\int dx\, |{\tilde{\psi}}_{\rm out}(x,y_m)|^2,
\end{align}
so that the normalized wave function of the output state can be written as
\begin{align}
  &
\psi_{\rm out}(x,y_m) =\frac{{\tilde{\psi}}_{\rm out}(x,y_m)}{\sqrt{P(y_m)}}.
\label{a016}
\end{align}}
{For the input squeezed coherent state of the target oscillator and the squeezed even/odd resource SCS, the probability density takes the explicit form 
\begin{align}
  &
\label{a0017}
    P(y_m)=\frac{1}{1+s e^{-r^2a^2}}\,\frac{\beta}{\sqrt{\pi}\,r}\, \exp\!{[-(y_m-Gx_0)^2\beta^2/r^2]} 
    \notag
  \\
  &
\times \Big(1+s e^{-a^2G^2/\gamma^2}\cos{[2a(y_m-Gx_0)\beta^2]}\Big),
\end{align}
where $\beta\equiv\delta/\gamma$.}
%
%

{As can be inferred from the expression (\ref{a0017}), in the case of an even ``perfect'' cat state at the gate output specified by the condition $y_m = Gx_0$, the probability density attains its maximum. In contrast, for the odd ``perfect'' cat state, the probability density may reach a local minimum, depending on the system parameters. At the same time, the depth of this dip decreases with increasing $a^2G^2/\gamma^2$ in the exponent, when the dominant contribution to $P(y_m)$ is determined by the prefactor outside parentheses in Eq. (\ref{a0017}). It should be noted that, as $ra$ increases, this prefactor becomes nearly identical for both the even and odd cat states, resulting in similar probability density behavior for these two quantum states: 
\begin{align}
\label{a18}
P(y_m)\approx \textcolor{black}{\frac{\beta}{\sqrt{\pi}\,r}}\exp\!{[-(y_m-Gx_0)^2\beta^2/r^2]}.
\end{align}
The same behavior is demonstrated by the probability density (\ref{a0017}) for the even cat state at small values of $a$ when $ra\ll 1$.}

\begin{figure*}
\centering
  \begin{tabular}{c @{\qquad} c @{\qquad} c}
    \includegraphics[width=0.5\linewidth]{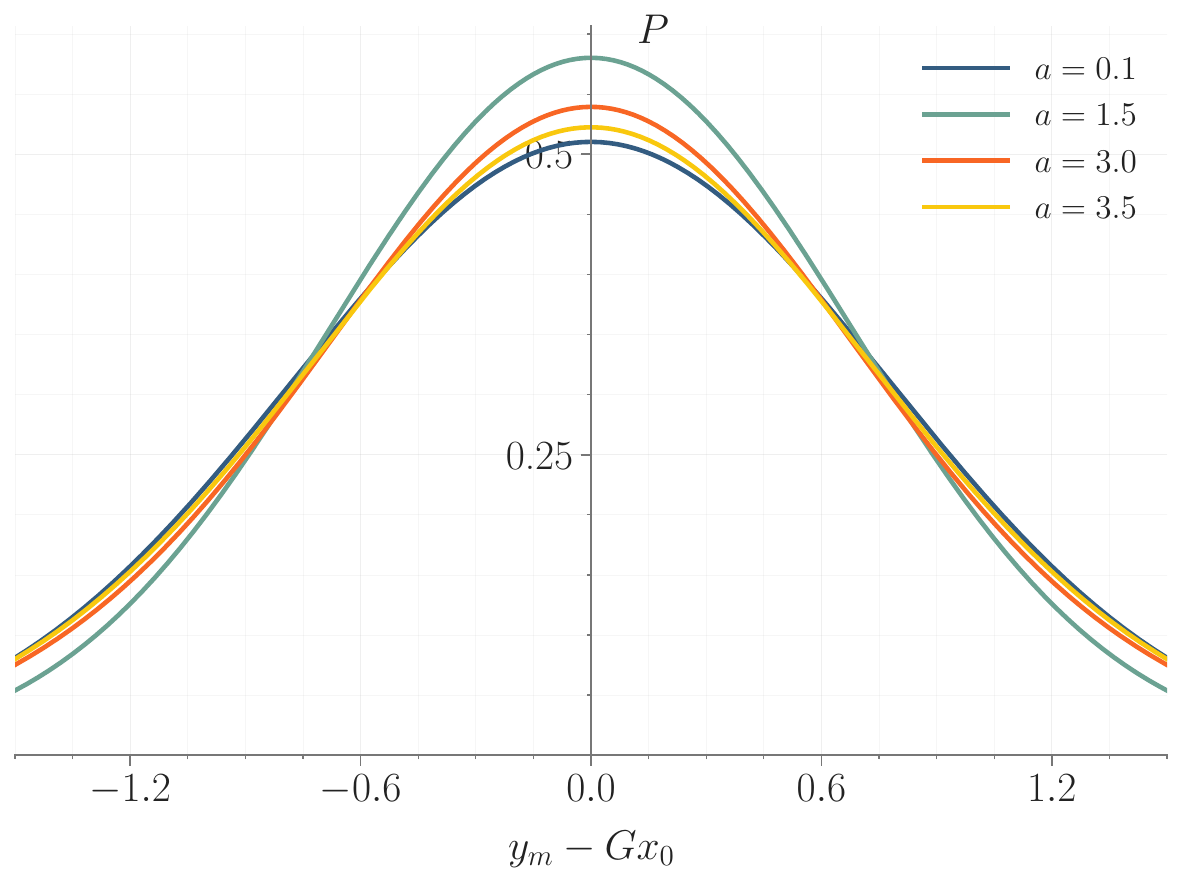} &
    \includegraphics[width=0.5\linewidth]{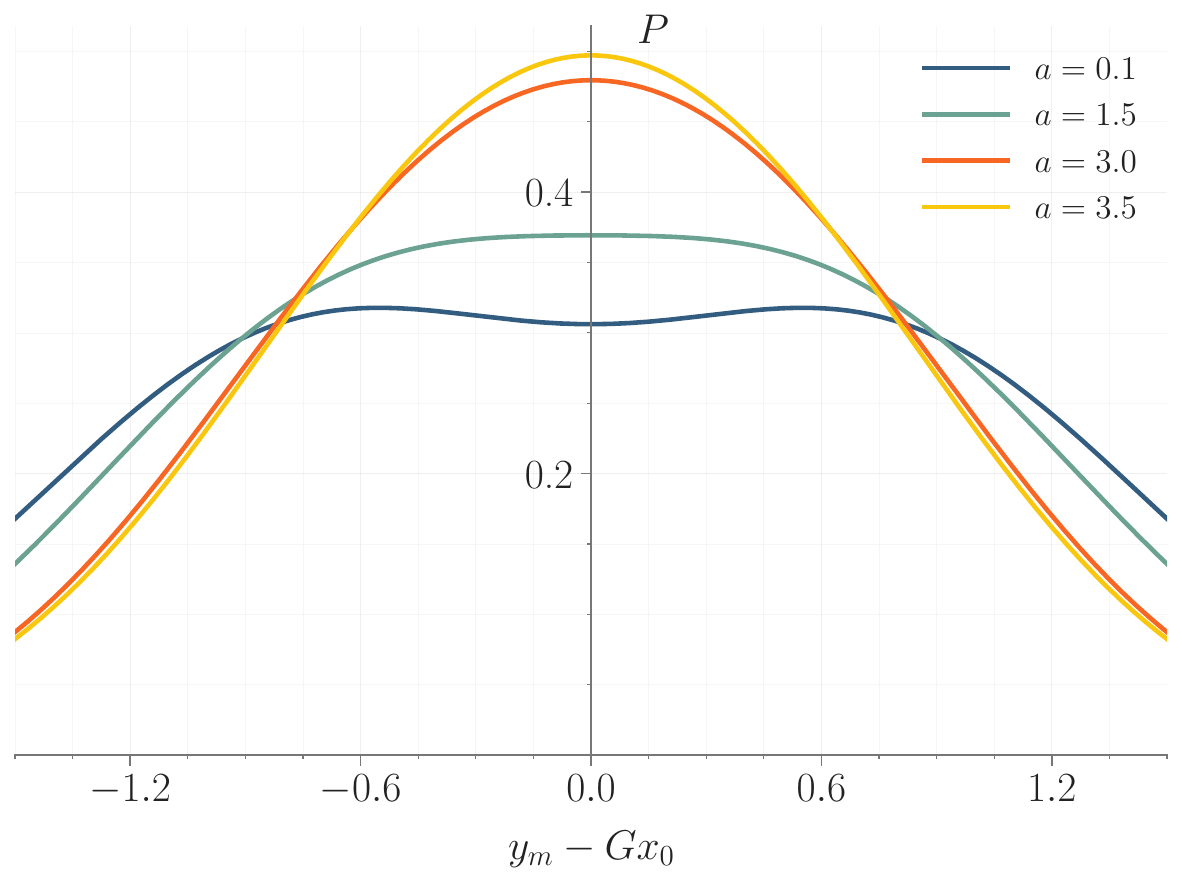}\\
    \small (a) & \small (b)
  \end{tabular}
  \caption{{A series of graphs depicting the probability density (\ref{a0017}) for the even (a) and odd (b) cat states as a function of $y_m - G x_0$ for various sizes $a=\{0.1, 1.5, 3.0, 3.5\}$ of the auxiliary SCS at the gate input, corresponding to the fixed parameter values: $r=0.5$, $\delta=2.0$, and $G=1.2$.}} 
  \label{figL2}
\end{figure*}

{The revealed trends are illustrated in Fig.~\ref{figL2} representing a series of graphs for the probability density (\ref{a0017}) for the even (Fig.~\ref{figL2} (a)) and odd (Fig.~\ref{figL2} (b)) cat states as a function of $y_m - G x_0$ for various sizes $a$ of the ancillary SCS at the gate input, corresponding to the fixed parameter values: $r=0.5$, $\delta=2.0$ and $G=1.2$.}

\begin{figure*}
\centering
  \begin{tabular}{c @{\qquad} c @{\qquad} c}
    \includegraphics[width=0.5\linewidth]{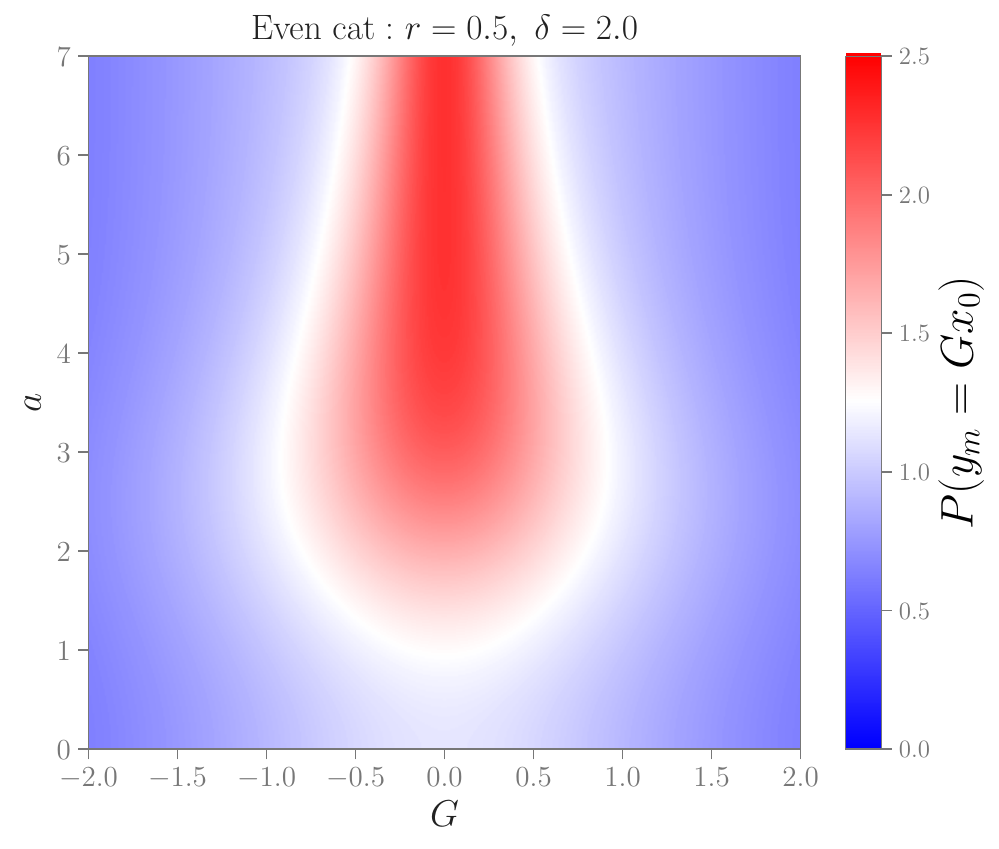} &
    \includegraphics[width=0.5\linewidth]{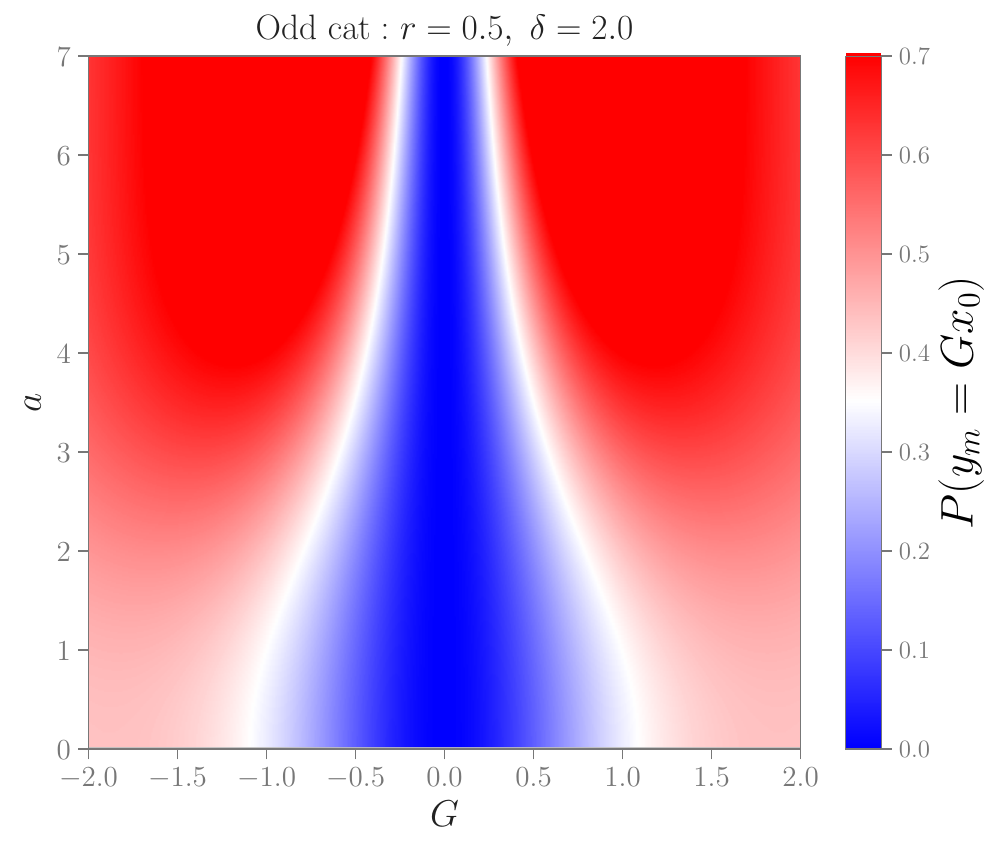}\\
    \small (a) & \small (b)
  \end{tabular}
  \caption{{Phase diagrams of the probability density for observing the outcome $y_m=Gx_0$, which heralds the appearance of an undistorted SCS at the gate output when the ancilla is in the even (a) or odd (b) SCS, as a function of $G$ and $a$ at fixed $r=0.5$ and $\delta=2.0$.  The color scale
    indicates the probability density ranging from $0$ (blue) to $1$
    (red).}} 
\label{figL44}
\end{figure*}

\begin{figure}[t!]
\centering
\includegraphics[width=1.0\columnwidth]{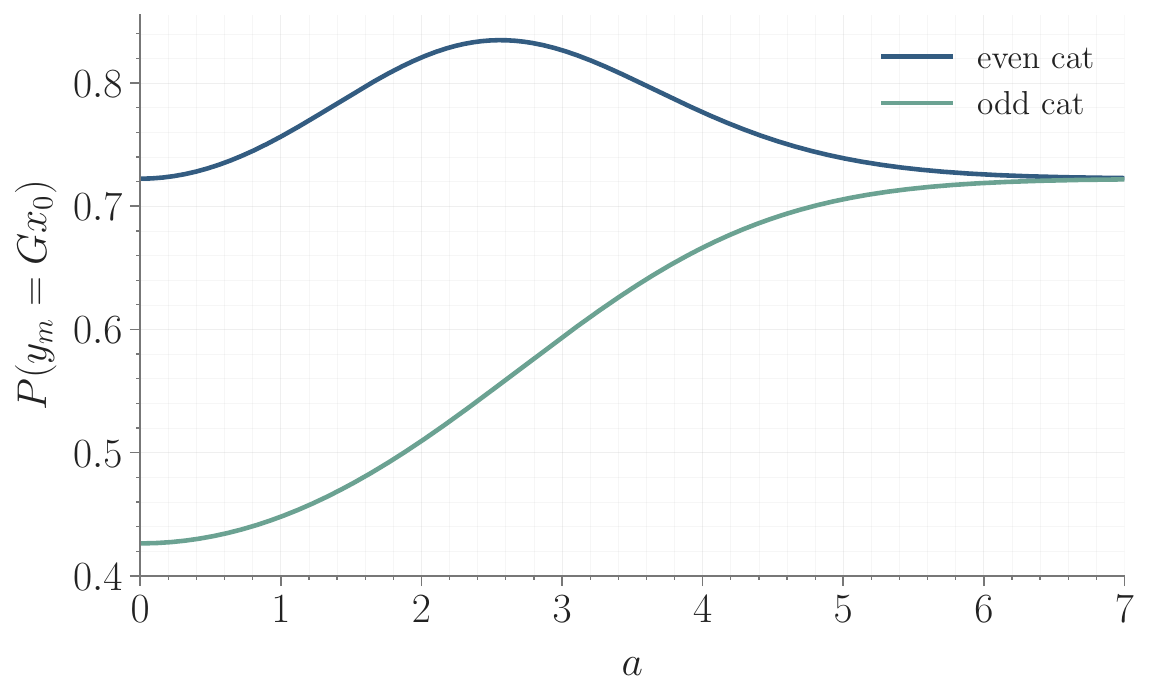}
\caption{{The probability density of the outcome $y_m=Gx_0$, which heralds the appearance of an undistorted SCS in the target output channel, depending on the size $a$ of the even/odd input ancilla SCS with fixed values of the parameters $r=0.5$, $\delta=2.0$, $G=1.2$.}}
\label{figL66}
\end{figure}

{The phase diagrams in Fig.~\ref{figL44}, along with the plots in Fig.~\ref{figL66}, provide detailed information on the probability density to observe an undistorted cat state at the gate output, when the measurement outcome is $y_m = Gx_0$.}


{From Figs.~\ref{figL44}, \ref{figL66} it is evident that, for the even SCS there exists a preferred size $a$ at which the probability density $P(y_m=Gx_0)$ attains its maximum. In contrast, for the odd SCS the probability density approaches a plateau as $a$ increases. Furthermore, values of the QND coupling strength $G$ for which a SCS with a smaller effective size ($|G|a$) than that at the gate input is generated, namely $|G|<1$, can yield a significantly higher probability density than those with $|G|>1$, for both parities of cat states.}

\subsection{Fidelity analysis}
\label{subsec:fidelity-analysis}

{Here we explore the conditions under which the state
  at the gate output defined by Eqs.~(\ref{a012}), (\ref{a016}) which
  is generally a distorted SCS most closely approximates the
  ``perfect'' even or odd squeezed SCS as the output target state
  given by equations (\ref{a11}), (\ref{a12}). For this purpose, we
  take the squeezed vacuum ($x_0=p_0=0$) as the input target state and
  estimate the ``proximity'' between the above mentioned two quantum
  states by considering the fidelity as the overlap between them,
\begin{align}
  &
\label{ab13}
F_{\rm cat}(y_m)= \left|\int dx \,\psi^{\rm cat}_{\rm out}(x)\,\psi^{*}_{\rm out}(x,y_m)\right|^2
\end{align}
depending on the measurement outcome $y_m$ and other relevant parameters.}

{As follows from the structure of expression
  (\ref{ab13}) and the explicit form of the wave functions
  (\ref{a012}) and (\ref{a12}), at $y_m\sim Gx_0$ the fidelity
  $F_{\rm cat}(y_m)$ reaches its largest values when the parities of
  the compared functions coincide and its smallest values when they
  are opposite.}

{The dominant maxima of $F_{\rm cat}(y_m)$ for same-sign wave functions are located approximately at
\begin{align}
\label{yy1}
y_ma\approx\pi k, \, k\in \mathbb{Z},
\end{align}
with the corresponding minima near 
\begin{align}
\label{yy2}
y_ma\approx\pi k+\pi/2, \, k\in \mathbb{Z}
\end{align}
for opposite-sign wave functions the roles are interchanged, so that
$y_ma\approx\pi k+\pi/2$ marks the approximate position of the maxima,
and $y_ma\approx\pi k$ that of the minima. These conditions originate
from the dominant interference between the two coherent-state branches
of the SCSs; the exact extrema of $F_{\rm cat}(y_m)$ may be slightly
shifted by the Gaussian envelope and the weaker oscillatory
modulations of the fidelity.}

For the functions (\ref{a012}) and (\ref{a12}) of the same parity, however,
$F_{\rm cat}(y_m)$ increases as $y_m\to Gx_0 = 0$, reaching unity.

\begin{figure*}
\centering
\includegraphics[width=2.1\columnwidth]{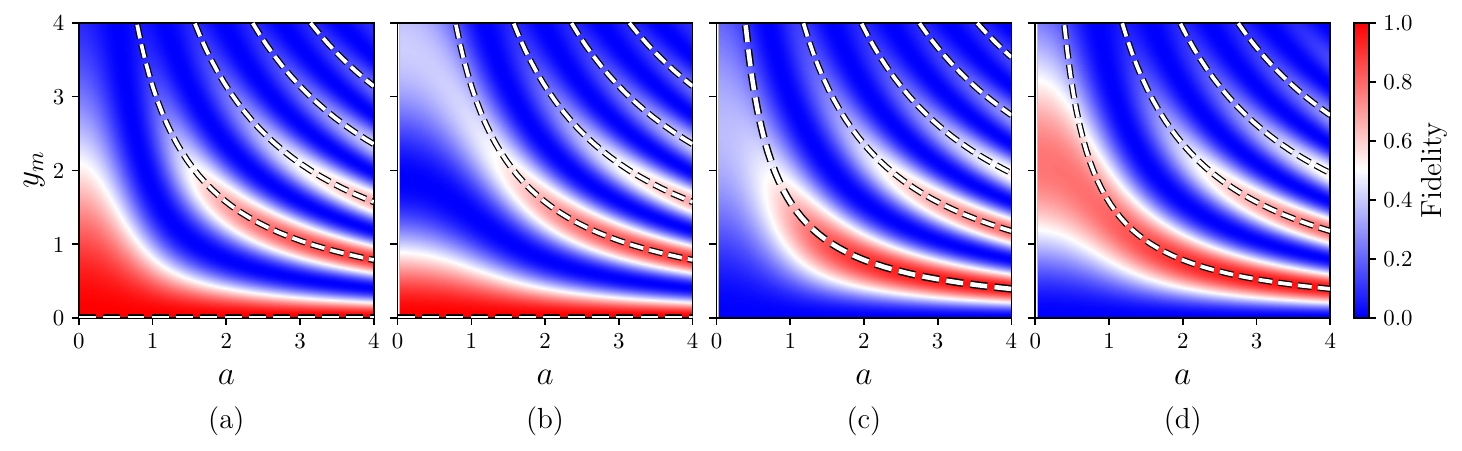}
\caption{{Phase diagrams of fidelity~(\ref{ab13}) between the
    ``perfect'' even or odd squeezed SCS (\ref{a12}) and the
    position-representation output wave function (\ref{a012}) obtained
    from an even or odd input SCS, for parameter values $G=1.2$,
    $\delta=0.5$, $r=1.5$; (a) each of the two wave functions is taken
    with the sign ``+'' (the ``even-even'' case); (b) each of the two
    wave functions is taken with the sign ``-'' (the ``odd-odd''
    case); (c) the wave function of the ``perfect'' cat state is taken
    with a ``-'' sign, while the exact output wave function is taken
    with a ``+'' sign (the ``odd-even'' case); (d) the wave function
    of the ``perfect'' cat state is taken with a ``+'' sign, while the
    exact output wave function is taken with a ``-'' sign (the
    ``even-odd'' case). Dashed lines indicate the maxima of
    $\cos(ay_m)$ for panels $(a)$ and $(b)$, and of $\sin(ay_m)$ for
    panels $(c)$ and $(d)$, which provide the dominant contributions
    to $F_{\rm cat}(y_m)$ in the regime of small $t$ at the Gaussian
    envelope $\exp(-\mu y_m^2/2r^4)$; see
    Eqs.~(\ref{ab15})-(\ref{ab18}). The color scale indicates
    the fidelity ranging from $0$ (blue) to $1$ (red).}}
\label{figL88}
\end{figure*}

{The described tendencies are confirmed by the phase diagrams for
  fidelity~(\ref{ab13}) between the undistorted even or odd squeezed
  cat state (\ref{a12}) and the output function (\ref{a012}) obtained
  from an even or odd input cat state shown in Fig.~\ref{figL88}, at
  the parameter values $G=1.2$, $\delta=0.5$, $r=1.5$. Panels $(a)$
  and $(b)$ illustrate the overlap of functions of identical parity
  (the same sign in the quantum superposition), while panels $(c)$ and
  $(d)$ depict the overlap of functions with opposite parity (opposite
  signs in the quantum superposition).}

%

{The nuances of behavior of each of the four phase diagrams can be
  comprehended through the explicit expressions for fidelity
  (\ref{ab13}):}

{\begin{align}
\label{ab15}
F^{++}_{\rm cat}(y_m)=\frac{e^{-\mu y^2_m/2r^4}(\cos{\!(ay_m)}+t\cos{\!(a\xi y_m)})^2}{(1+t)(1+t\cos{\!(2a\xi y_m)})},
\end{align}
}

{
\begin{align}
\label{ab16}
F^{--}_{\rm cat}(y_m)=\frac{e^{-\mu y^2_m/2r^4}(\cos{\!(ay_m)}-t\cos{\!(a\xi y_m)})^2}{(1-t)(1-t\cos{\!(2a\xi y_m)})},
\end{align}}

{
\begin{align}
\label{ab17}
F^{+-}_{\rm cat}(y_m)= \frac{e^{-\mu y^2_m/2r^4}(\sin{\!(ay_m)}+t\sin{\!(a\xi y_m)})^2}{(1+t)(1-t\cos{\!(2a\xi y_m)})},
\end{align}}

{
\begin{align}
\label{ab18}
F^{-+}_{\rm cat}(y_m)=\frac{e^{-\mu y^2_m/2r^4}(\sin{\!(ay_m)}-t\sin{\!(a\xi y_m)})^2}{(1-t)(1+t\cos{\!(2a\xi y_m)})},
\end{align}
where the superscripts specify the sign within the quantum
superposition for the first state (the ``perfect'' squeezed SCS) and
the second state (the exact solution at the gate output);
$\mu\!\equiv G^2r^2/(\delta^2r^2+G^2)$, $\xi\!=\mu\delta^2/G^2$,
$t\!=\exp{(-\mu a^2)}$. For small values of $t$, the principal
contributions to the fidelity $F_{\rm cat}(y_m)$ at the exponential
term $\exp{(-\mu y^2_m/2r^4)}$ are $\cos{\!(ay_m)}$ for (\ref{ab15})
and (\ref{ab16}), and $\sin{\!(ay_m)}$ for (\ref{ab17}) and
(\ref{ab18}). The dominant extrema associated with the conditions
(\ref{yy1}) and (\ref{yy2}) are marked by the dashed lines in
Fig.~\ref{figL88}.}



\subsection{Generation of squeezed cats from vacuum and
  small-amplitude cat-states}
\label{subsec:gener-sq-scs}

{From a practical implementation perspective, it seems more attractive
  when unsqueezed SCSs of small size are used to generate a squeezed
  large-amplitude SCS. Thus, in this section, we consider the
  generation of a squeezed SCS from an unsqueezed vacuum state,
  $\delta=1$, $x_0=p_0=0$, based on the ancillary unsqueezed SCS,
  $r=1$, of any parity ($s=\pm1$)
\begin{align}
  \label{a0014}
  &
\!{\psi}_{\rm anc}(x)\! =\!\frac{1}{\sqrt{2\pi^{1/2}(1\!+s
  e^{-a^2})}}\!\Bigl(e^{-(x-a)^2/2}
  \notag
  \\
  &
  +s e^{-(x+a)^2/2}\Bigr).
\end{align}} 

{Using Eq. (\ref{a012}), we obtain the expression for the unnormalized
  output wave function of the squeezed even or odd SCS
\begin{align}
\label{a014}
\!{\tilde{\psi}}_{\rm out}(x)\! \sim\!\frac{e^{-(G^2+1)x^2/2}}{\sqrt{2\pi(1\!+s e^{-a^2})}}\!\Big(e^{ixGa}\!+s e^{-ixGa}\Big),
\end{align}
when the measurement outcome is $y_m=Gx_0=0$.}

{From Eq.~(\ref{a014}) we conclude that, in the present context
  $\gamma=\sqrt{1+G^2}$, our scheme conditionally generates a squeezed
  SCS with quadrature squeezing factor $\gamma^{-1}$ and component
  separation $2|G|a$ from the vacuum state of the target oscillator
  using an ancillary Schr{\"o}dinger kitten or cat state. In
  principle, increasing $G$ enlarges the effective size of the
  output SCS, although in practice this simultaneously increases
  squeezing, reduces the finite-window success probability
  (see Section~\ref{subsec:prob_dens_k}), and is limited by experimentally accessible QND strengths.}

\begin{figure*}
\centering
  \begin{tabular}{c @{\qquad} c @{\qquad} c}
    \includegraphics[width=0.5\linewidth]{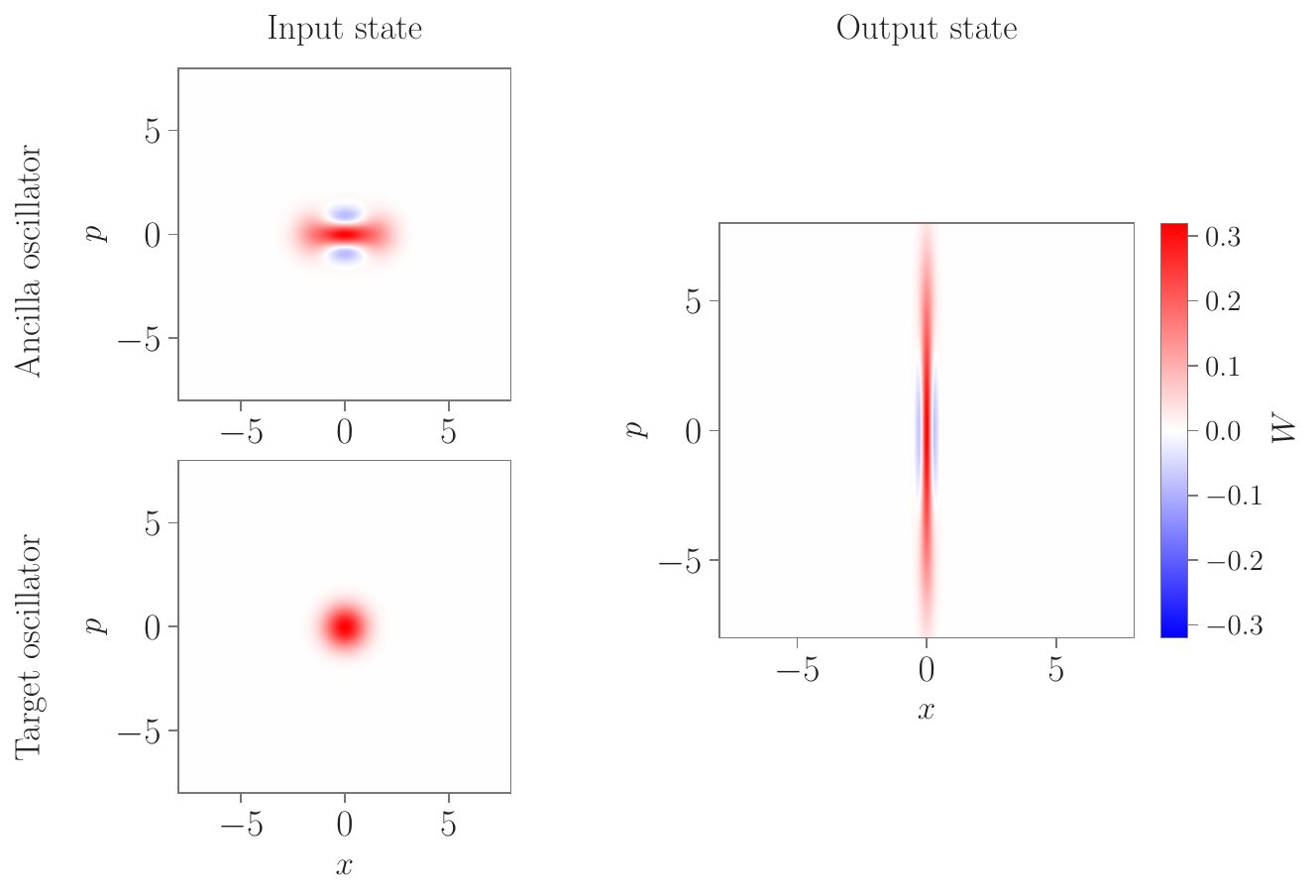} &
    \includegraphics[width=0.5\linewidth]{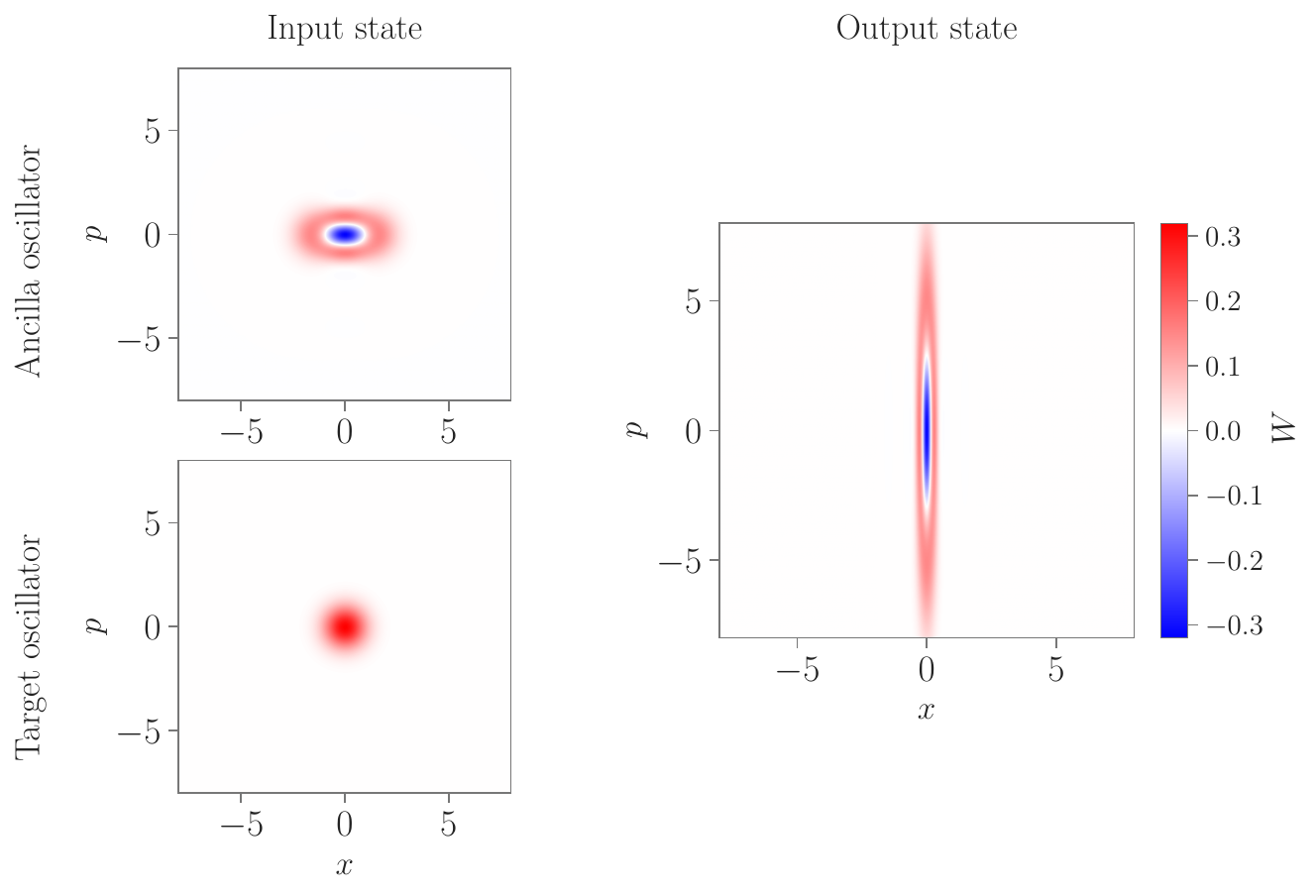}\\
        \small (a) & \small (b)
    \end{tabular}
    \caption{{Wigner functions of the input quantum states — the
        target vacuum state and the ancillary kitten state with the
        size $a=\sqrt{2}$, which corresponds to the coherent
        amplitude $\alpha=1$, — and of the output state - the squeezed
        even (a) and odd (b) SCSs, obtained for the gate parameters
        $\delta=r=1$ and $G=3$. The effective size of the prepared
        squeezed SCS is $|G|a\approx 4.243$, while the quadrature
        squeezing factor is
        $\gamma^{-1}=(1+G^2)^{-1/2}\approx 0.316$.}}
\label{figB4}
\end{figure*}

{Figure~\ref{figB4} shows the Wigner functions of the initial
  unsqueezed states—the target vacuum state and the ancillary kitten
  state with the size $a=\sqrt{2}$ (which corresponds to the amplitude $\alpha=1$) — as well as those of the output
  squeezed even (a) and odd (b) SCSs, obtained for a QND interaction
  strength $G=3$. The prepared squeezed cat effective size is
  $|G|a\approx 4.243$, while the quadrature squeezing factor is
  $\gamma^{-1}=(1+G^2)^{-1/2}\approx 0.316$.}


\section{Iterative growth of squeezed cats}
\label{section_III}

{Here we describe a protocol for amplifying optical SCSs by the
  iterative growth of their size and squeezing based on the gate
  represented in Section~\ref{section_II} which is applied in an
  iterative way~\cite{Sychev_2017, Etesse_2015, Laghaout_2013}. We
  assume that in each iteration, the protocol consists of: 1) a QND
  entangling operation $\hat C_Z$ with fixed coupling constant $G$; 2)
  a projective homodyne measurement of the ancilla momentum; 3)
  postselection on the measurement outcome; 4) a $-\pi/2$ phase-space
  rotation before the next iteration. Crucially, the target input
  state in every step is the vacuum, while the ancilla state at
  iteration $k$ is the rotated conditional output state obtained at
  iteration $k-1$. This measurement-induced cat-state amplification
  scheme allows one, in principle, to increase the cat amplitude
  iteratively, within practical limits set by the available QND
  strength.}

{Suppose that there are states of unsqueezed vacuum and unsqueezed SCS
  described by Eq.~(\ref{a3}) at $r\equiv r_{(0)}=1$, and the
  operation $\hat C_Z$ is characterized by a fixed value of the QND
  coupling coefficient $G$ such that $|G|\gtrsim 1$.}

{The first application of the gate under consideration at the
  measurement outcome $y_m=Gx_0=0$ of the ancillary oscillator
  momentum heralds the appearance of an undistorted SCS (\ref{a014})
  at the gate target output. This state is oriented along the $p$-axis
  in phase space squeezed by the factor $\sqrt{G^2+1}\equiv r_{(1)}$
  (the inverse quadrature squeezing factor), and has a distance
  between components equal to $2|G|a$,
\begin{gather} 
\psi^{\mathrm{cat(1)}}_{\mathrm{out}}(x)\sim e^{-(G^2+1)x^2/2}\Big(e^{ixGa}+s e^{-ixGa}\Big),
\label{a20}
\end{gather}
} {By rotating in phase space the squeezed SCS (\ref{a20}) by the
  angle $-\pi/2$, we arrive at the state that we will use as the
  initial ancillary state for the second iteration,
\begin{gather} 
\psi^{(2)}_{in}(x_2)\sim \nn\\
\exp\{-r^{-2}_{(1)}(x_2-Ga)^2/2\}+s \exp\{-r^{-2}_{(1)}(x_2+Ga)^2/2\},
\label{a21}
\end{gather}
and then, after applying the gate twice, we get the output undistorted squeezed SCS
\begin{align}
\label{a22}
\psi^{\mathrm{cat(2)}}_{\mathrm{out}}(x)\sim e^{-r^2_{(2)}x^2/2}\Big(e^{ixG^2a}+s e^{-ixG^2a}\Big),
\end{align}
characterized by the inverse quadrature squeezing factor $r_{(2)}\equiv\sqrt{1+G^2(1+G^2)}$ and the effective size $G^2a$.}

{The iterative procedure performed $k$-times provides the following
  heralded undistorted squeezed SCS at the output of the circuit
\begin{align}
\label{a23}
\psi^{\mathrm{cat(k)}}_{\mathrm{out}}(x)\sim e^{-r^2_{(k)}x^2/2}\Big(e^{ixG^ka}+s e^{-ixG^ka}\Big),
\end{align}
having the inverse quadrature squeezing factor
$r_{(k)}=\sqrt{\Sigma^{k}_{j=0}(G^2)^j}$ and the effective size
$|G|^k a$. The normalization factor of the ``perfect'' squeezed SCS of
Eq.~(\ref{a23}), introduced in the same way as $\mathcal{N}$ of
Eq.~(\ref{a11}), is
\begin{equation}
\mathcal{N}^{(k)}_{\mathrm{cat}}
= \frac{2\sqrt{\pi}}{\sqrt{A_{k}}}\,(1 + s\,\mathcal{V}_k),
\label{g7}
\end{equation}
with $A_{k}$ and $\mathcal{V}_k$ as defined in Appendix~\ref{app1}.}

\begin{figure*}
\centering
\includegraphics[width=2.2\columnwidth]{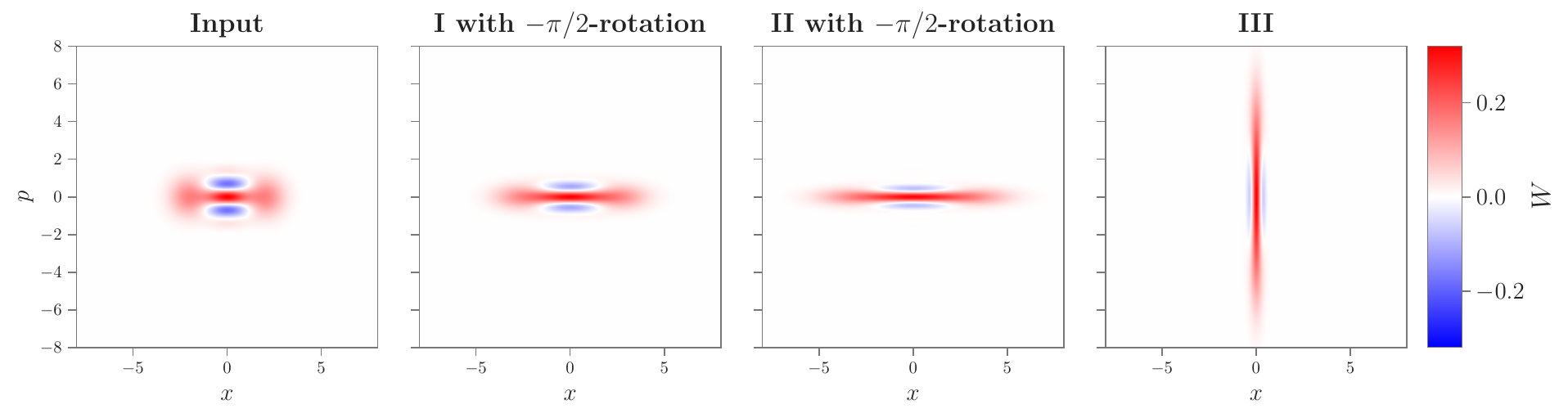}
\caption{{Scheme for iterative conditional cat-state engineering. The
    Wigner functions correspond to three successive gate iterations
    for $r_{(0)}=1$ $G=1.2$, $a=2.0$. The inverse quadrature squeezing
    factors at each step are $r_{(1)}=\sqrt{1+G^2}\approx 1.56$,
    $r_{(2)}=\sqrt{1+G^2+G^4}\approx 2.12$, and
    $r_{(3)}=\sqrt{1+G^2+G^4+G^6}\approx2.74$. The corresponding
    effective sizes of SCSs are $|G|a=2.40$, $G^2a=2.88$, and
    $|G|^3a\approx3.46$, respectively.}}
\label{figA1555}
\end{figure*}
%


\subsection{Total success probability in multi-step cat-state amplification protocol}
\label{subsec:prob_dens_k}
 
{Since in actual experiments the homodyne outcome is not post-selected
  to a single value $y_m=Gx_0$, but can only be resolved with a finite
  precision determined by the measurement window of the apparatus, it
  is appropriate to consider a mixed output state which arises when
  the measured ancilla momentum $y_m$ falls within a symmetric
  acceptance interval $y_m \in [-d/2,d/2]$ centered at $y_m=Gx_0=0$
  for $x_0=0$.}

{For a given input state $\rho_T$, the unnormalized conditional output
  state after the $\hat C_Z$ application followed by a homodyne
  measurement of the ancilla momentum with outcome $y$ (but before
  rotation) is
\begin{gather} 
\tilde{\rho}_T(y)=
\operatorname{Tr}_{A}
\left[
(\mathbb{I}_T \otimes |y\rangle\langle y|)
\hat C_Z
(\rho_T \otimes \rho_A)
\hat C_Z^\dagger
\right].
\label{x1}
\end{gather}
Here $|y\rangle\langle y|$ is the projector corresponding to the
homodyne outcome $y$; $\hat C_Z = e^{iG \hat x_T \hat x_A}$ is the QND
entangling operation; the indices $T$ and $A$ indicate the target and
measured ancillary modes, respectively.}

{The probability density of outcome $y$ is obtained by taking the
  trace of the unnormalized state~(\ref{x1}):
\begin{gather} 
P(y) = \operatorname{Tr}_T[\tilde{\rho}_T(y)],
\label{x2}
\end{gather}
and the normalized conditional state is $\rho_T(y) = \tilde{\rho}_T(y)/P(y)$.}

{For the first iteration the input target state is vacuum,
  $\rho_T\equiv\rho^{(0)}_T = |0\rangle\langle 0|$, and the input
  ancillary state is
  $\rho^{(0)}_{A_1}=|\psi_{\rm anc}\rangle\langle\psi_{\rm anc}|$ with
  coordinate wave function~(\ref{a0014}). If outcomes are accepted
  only when $y_1 \in [-d/2,d/2]$, the success probability of the first
  iteration is
\begin{gather}
P_{(1)}(d)=\int^{d/2}_{-d/2} dy_1\; P_1(y_1)
\label{x3}
\end{gather}
where the probability density $P_1(y_1)$ is given by Eq.~(\ref{a0017}).}

{At each subsequent iteration $j\ge 2$, the target input is always a
  vacuum state, while the ancilla is the $-\pi/2$-rotated conditional
  output of the preceding step. Since the conditional output of step
  $j{-}1$ depends on the entire measurement history
  $\mathbf{y}_{<j} \equiv (y_1,\dots,y_{j-1})\in\mathbb{R}^{k-1}$, the
  ancilla state entering step $j$ inherits this dependence:
\begin{equation}
\rho_{A_j}^{(j-1)}(\mathbf{y}_{<j})
= \frac{R\,\tilde{\rho}_T^{(j-1)}(\mathbf{y}_{<j})\,R^\dagger}
{P_{j-1}(y_{j-1} \mid \mathbf{y}_{<j-1})},
\label{eq:rhoAj}
\end{equation}
where $R \equiv R(-\pi/2) = e^{i\pi \hat{a}^\dagger \hat{a}/2}$
denotes the phase-space rotation operator by $-\pi/2$ acting on
quadratures according to $R^\dagger \hat{x} R = -\hat{p}$ and
$R^\dagger \hat{p} R = \hat{x}$. The QND entangling operation
$\hat C_{\!Z}$ followed by homodyne projection onto $|y_j\rangle$ then
yields the unnormalized conditional target state
\begin{align}
  &
\tilde{\rho}_T^{(j)}(\mathbf{y}_{j})
\!= \!\mathrm{Tr}_{A_j}\!\Bigl\{
    (\mathbb{I}_T \otimes |y_j\rangle\langle y_j|) \notag
  \\
  &
    \times
\hat C_{\!Z}\bigl(\rho_T^{(0)} \otimes \rho_{A_j}^{(j-1)}(\mathbf{y}_{<j})\bigr)\hat C_{\!Z}^\dagger
\Bigr\}\!,
\label{eq:rhoj}
\end{align}
and the corresponding conditional probability density of the $j$-th outcome reads
\begin{equation}
P_j(y_j \mid \mathbf{y}_{<j}) = \mathrm{Tr}_T\bigl[\tilde{\rho}_T^{(j)}(\mathbf{y}_{j})\bigr],
\label{eq:Pj_cond}
\end{equation}
where $\mathbf{y}_{j} \equiv (y_1,\dots,y_{j})\in\mathbb{R}^k$. The
nested structure of Eqs.~(\ref{eq:rhoAj}), (\ref{eq:rhoj}) makes the
dependence on the full measurement history explicit: the ancilla state
$\rho_{A_j}^{(j-1)}$ at step $j$ is determined by
$\tilde{\rho}_T^{(j-1)}$, which in turn depends on
$\rho_{A_{j-1}}^{(j-2)}$, and so on down to the initial cat state at
step $j=1$. Consequently, the measurement outcomes at successive steps
are not statistically independent.}

{Consider a sequential protocol consisting of $k$ homodyne
  measurements with outcomes $\mathbf{y}_k$. Success is defined as all
  outcomes falling inside the symmetric window $|y_j|\le d/2$. The
  joint probability density of the full outcome vector $\mathbf{y}_k$
  is related to the conditional densities~\eqref{eq:Pj_cond} by the
  probability chain rule,
\begin{equation}
P(\mathbf{y}_k)
= P_1(y_1)\prod_{j=2}^{k} P_j(y_j \mid \mathbf{y}_{<j}),
\label{eq:chain_rule}
\end{equation}
which is a general identity valid for any sequential measurement
process. However, direct evaluation of Eq.~\eqref{eq:chain_rule} by
iterating the quantum-state
recursion~\eqref{eq:rhoAj}--\eqref{eq:Pj_cond} step by step would be
computationally prohibitive for large~$k$, since at each step the full
conditional density matrix must be propagated as a function of all
preceding outcomes. In Appendix~A, we circumvent this difficulty by
adopting a different strategy: we decompose the initial cat-state
density matrix into its two coherent-state branches and derive a
closed-form recursion [Eq.~\eqref{S3}] for the branch wave functions
that can be solved explicitly [Eq.~\eqref{S6}]. The joint density is
then reconstructed directly from the overlaps of these branch wave
functions after all $k$ steps, yielding a closed-form expression
[Eq.~\eqref{S11}] without the need to evaluate the chain-rule
product.}

{The total probability of success after $k$ iterations is
\begin{align}
  &
\label{U1}
P_{(k)}(d)= \int_{\mathcal{D}_d} P(\mathbf{y}_k)\,d^k\!{\mathbf y_k},
\end{align}
where $\mathcal{D}_d \equiv [-d/2,\,d/2]^k$ and
$d^k\mathbf{y}_k \equiv dy_1\cdots dy_k$. Substituting the closed-form
expression for the joint probability density derived in
Appendix~\ref{app1} [Eq.~(\ref{S11})], the total success probability
can be written as
\begin{align}
\label{R26}
P_{(k)}(d) = \frac{J_k^{(0)}(d)+s\,e^{-a^2 G^{2k}/A_k}\;\mathrm{Re}\,J_k^{(1)}(d)}{1+s\,e^{-a^2}},
\end{align}
where the quantities entering this expression are defined as follows:
$s=+1$ for the even and $s=-1$ for the odd initial ancillary cat
state; $a=\sqrt{2}\alpha$ is the size parameter of the
coherent state $|\alpha\rangle$ ;
$A_k\equiv r_{(k)}^2=\sum_{l=0}^k G^{2l}$ is the squared inverse
quadrature squeezing factor of the output cat state after $k$
iterations. The Gaussian integral and the interference integral are
\begin{align}
  &
J_k^{(0)}(d)\! = \!\!\int_{\mathcal{D}_d} \!\!\!W_k(\mathbf
y_k)\,d^k\!{\mathbf y_k},
\notag
  \\
  &
J_k^{(1)}(d)\! = \!\!\int_{\mathcal{D}_d}\!\!\!W_k(\mathbf y_k)\,e^{i\Psi_k(\mathbf y_k)}\,d^k\!{\mathbf y_k}. 
\label{z8main}
\end{align}
The multivariate Gaussian envelope is
\begin{gather}
W_k(\mathbf y_k)=\dfrac{1}{\pi^{k/2}\sqrt{A_k}}\exp\!\left(\dfrac{E_k^2}{A_k}-F_k\right), 
\label{z88}
\end{gather}
where $E_k=\sum_{l=1}^k(-G)^l\phi_l$, $F_k=\sum_{l=1}^k\phi_l^2$, and
the auxiliary shifts obey the recursion
$\phi_l=y_{k+1-l}-G\,\phi_{l-1}$, $\;\phi_0=0$. The interference
global phase, which is linear in the outcomes, reads
\begin{gather}
\Psi_k(\mathbf{y}_k)=2a\!\left[\phi_k-\dfrac{(-G)^k E_k}{A_k}\right].
\label{z888}
\end{gather}}
{The denominator $1+s\,e^{-a^2}$ is the global normalization constant $\mathcal{N}_k$ which is proven in Appendix~\ref{app2} to be independent of the iteration number $k$.}

{To evaluate the $k$-dimensional integrals $J_k^{(0)}$ and $J_k^{(1)}$
  efficiently, in Appendix~B we factorize the Gaussian envelope $W_k$
  into a product of conditional single-variable Gaussians (a
  Cholesky-type decomposition) with inverse-width parameters
  $\eta_j = r_{(j-1)}/r_{(j)}$, conditional means $\mu_1 = 0$,
  $\mu_{j+1} = G\, r_{(j-1)}^2 (y_j - \mu_j)/r_{(j)}^2$, and
  interference frequencies $\delta_j =
  2a\,(-G)^{j-1}/r_{(j)}^2$. Although this factorization is
  structurally reminiscent of the chain rule~\eqref{eq:chain_rule}, it
  is a purely mathematical decomposition of the already known
  quadratic form in the exponent of $W_k$, not a re-derivation of the
  joint density from conditional quantum states. The resulting product
  form reduces the $k$-dimensional integration to a nested sequence of
  one-dimensional integrals, with the innermost building
  block expressible in terms of the error function of complex
  argument; see Appendix~\ref{app2} for explicit formulae.}

\begin{figure}
\centering
\includegraphics[width=1.0\linewidth]{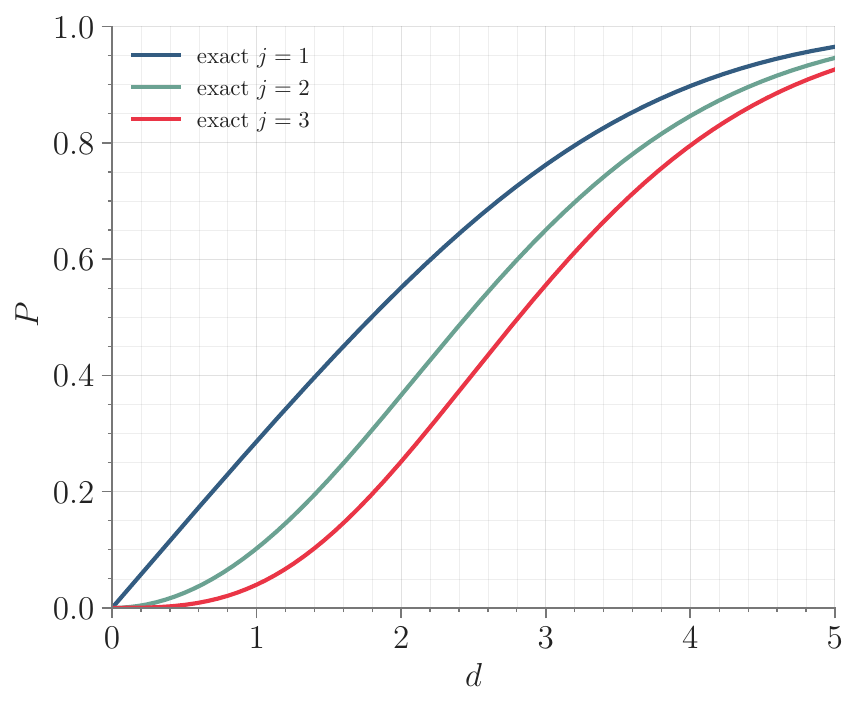}
\caption{Total success probability $P_{(k)}(d)$ within the acceptance
  window $[-d/2, d/2]$ for an odd initial ancillary SCS after $k$
  iterations ($k=1, 2, 3$) with the parameter values $a=\sqrt{2}$,
  $G=1.2$, $\delta=1$, $r_{(0)}=1$. The graphs are based on the
  expression (\ref{R26}).}
\label{F15}
\end{figure}

{Figures~\ref{F105}, \ref{F15} present the total success probability
  $P_{(k)}(d)$ that all $k$ measurement outcomes fall within the
  acceptance interval $[-d/2, d/2]$ for an even/odd initial ancillary
  SCS after $k=1, 2, 3$ iterations. The curves are computed from
  Eq.~(\ref{R26}) with $a=\sqrt{2}$, $G=1.2$, $\delta=1$,
  $r_{(0)}=1$.}
\begin{figure}
\centering
\includegraphics[width=1.0\linewidth]{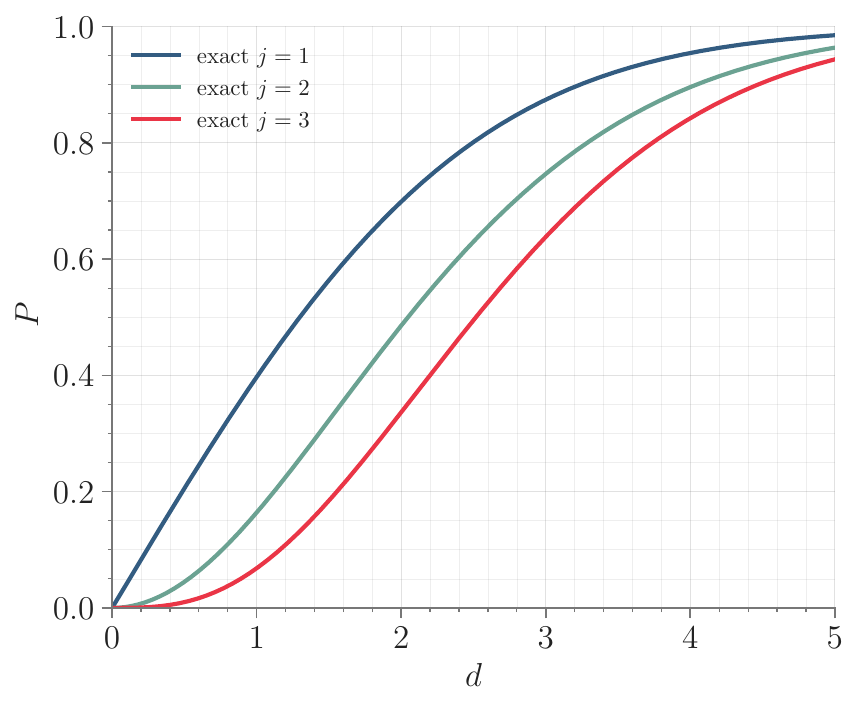}
\caption{Total success probability $P_{(k)}(d)$ within the acceptance
  window $[-d/2, d/2]$ for an even initial ancillary SCS after $k$
  iterations ($k=1, 2, 3$) with the parameter values $a=\sqrt{2}$,
  $G=1.2$, $\delta=1$, $r_{(0)}=1$. The graphs are based on the
  expression (\ref{R26}).}
\label{F105}
\end{figure}
%

\begin{figure*}[t!]
\centering
  \begin{tabular}{c @{\qquad} c @{\qquad} c}
    \includegraphics[width=0.33\linewidth]{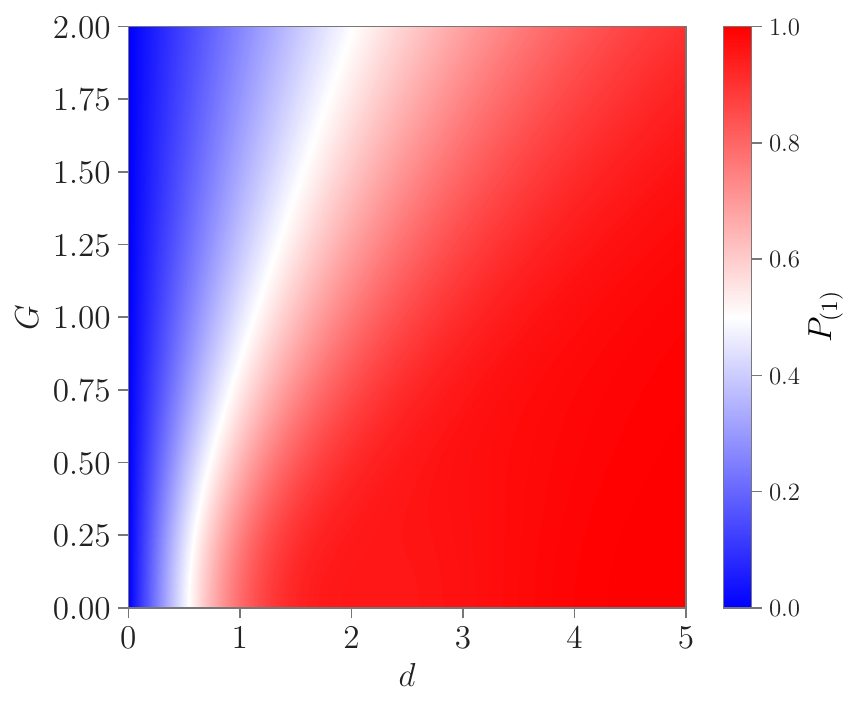} &
    \includegraphics[width=0.33\linewidth]{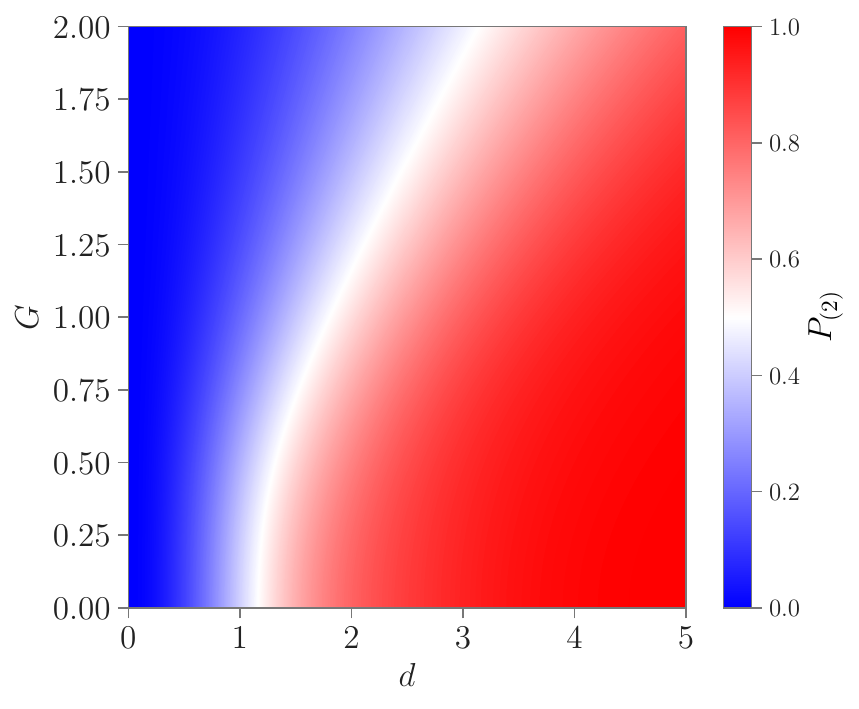} &
    \includegraphics[width=0.33\linewidth]{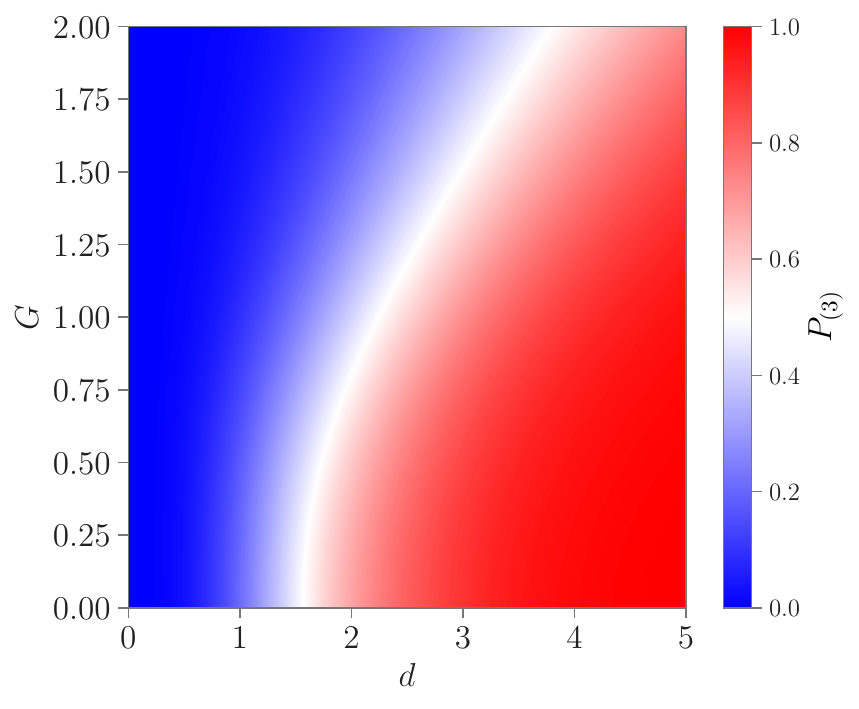} \\
    \small (a) $\,\,k=1$  & \small (b) $\,\,k=2$ & \small (c) $\,\,k=3$
  \end{tabular}
  \caption{Phase diagrams of the total success probability
    $P_{(k)}(d,G)$ within the acceptance window $[-d/2,d/2]$ for an
    even initial ancillary SCS for the number of iterations
    $k=1, 2, 3$ of the cat-state growth protocol with the parameter values
    $a=\sqrt{2}$, $\delta=1$, $r_{(0)}=1$. The color scale indicates
    the total success probability ranging from $0$ (blue) to $1$ (red).}
\label{figF16}
\end{figure*}

\begin{figure*}[t!]
\centering
  \begin{tabular}{c @{\qquad} c @{\qquad} c}
    \includegraphics[width=0.33\linewidth]{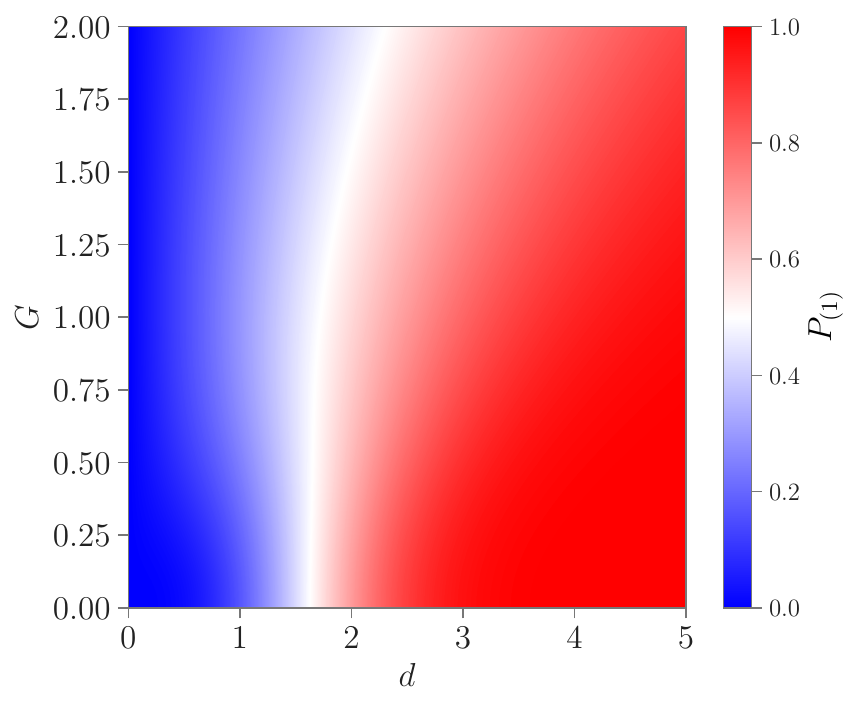} &
    \includegraphics[width=0.33\linewidth]{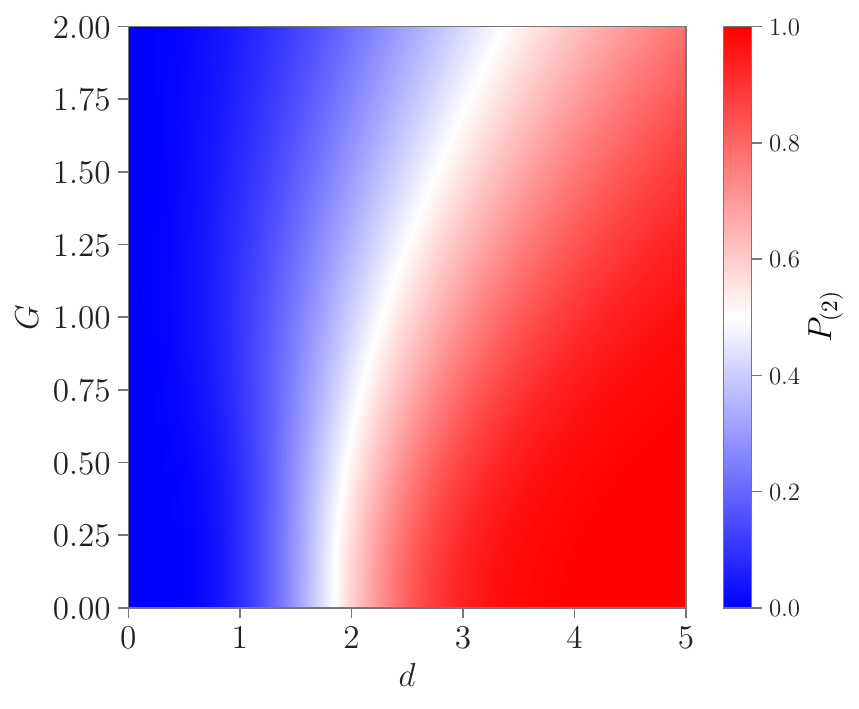} &
    \includegraphics[width=0.33\linewidth]{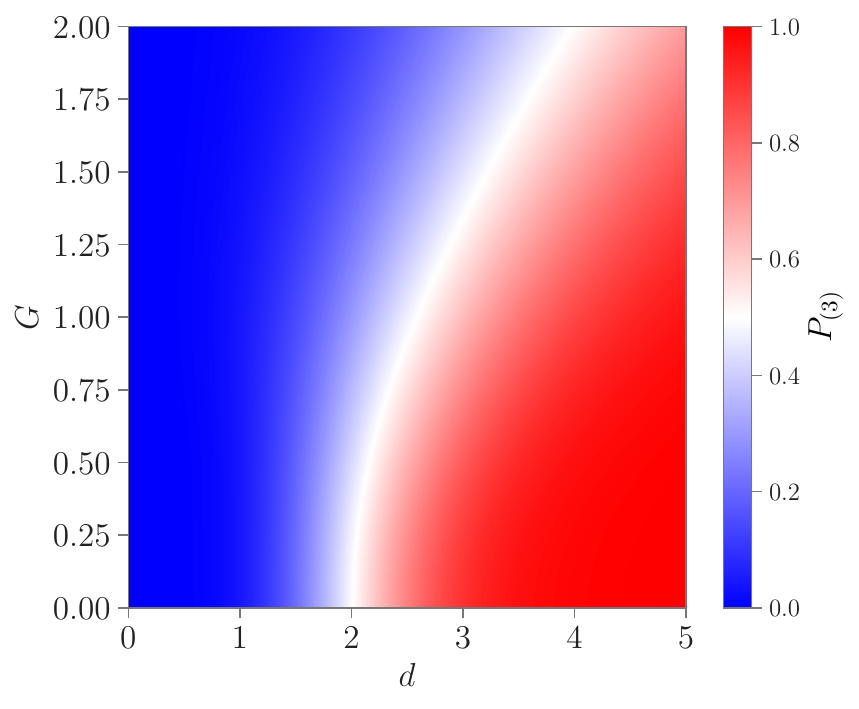} \\
    \small (a) $\,\,k=1$  & \small (b) $\,\,k=2$ & \small (c) $\,\,k=3$
  \end{tabular}
  \caption{Phase diagrams of the total success probability
    $P_{(k)}(d,G)$ within the acceptance window $[-d/2,d/2]$ for an
    odd initial ancillary SCS for the number of iterations $k=1, 2, 3$
    of the cat-state growth protocol with the parameter values
    $a=\sqrt{2}$, $\delta=1$, $r_{(0)}=1$. The color scale indicates
    $P_{(k)}$ ranging from $0$ (blue) to $1$ (red).}
\label{figF17}
\end{figure*}

{Figures~\ref{figF16} and \ref{figF17} show the phase diagrams of the
  total success probability $P_{(k)}$ as a function of the
  acceptance-window width $d$ and the QND coupling strength $G$ for
  even and odd initial ancillary SCS, respectively, with
  $a = \sqrt{2}$, $\delta=1$, $r_{(0)}=1$. Panels~(a)--(c) correspond
  to $k=1, 2, 3$ iterations. For each $k$, the total success
  probability decreases with increasing $G$ and decreasing $d$,
  regardless of the initial cat-state parity. As $k$ grows, the region
  of high $P_{(k)}$ shrinks along both axes, reflecting the cumulative
  impact of finite homodyne resolution: measurement-induced
  imperfections accumulate with each iteration, making the protocol
  progressively more sensitive to the acceptance-window settings.}



\subsection{Mixed-state fidelity after iterations}
\label{subsec:fmix_k_exact}

{In the $k$-step protocol, the conditional output at each step depends
  on the full measurement history $\mathbf{y}_{<j}$ [see
  Eq.~(\ref{eq:rhoj})], so a finite acceptance window
  $y_j \in [-d/2, d/2]$ turns the post-selected output into a
  $k$-dimensional classical mixture of conditional states. Our figure
  of merit for this mixture is the fidelity with respect to the
  undistorted squeezed SCS
  $|\psi^{\mathrm{cat}(k)}_{\mathrm{out}}\rangle$ of effective
  size $|G|^k a$ and the inverse quadrature squeezing factor
  $r_{(k)}$, whose wave function is given by Eq.~(\ref{a23}):
\begin{equation}
    F^{\mathrm{mix}}_{(k)}(d)
    \equiv \langle \psi^{\mathrm{cat}(k)}_{\mathrm{out}} | \rho^{\mathrm{mix}}_{(k)}(d) | \psi^{\mathrm{cat}(k)}_{\mathrm{out}} \rangle,
    \label{eq:F_mix_def}
\end{equation}
where $\rho^{\mathrm{mix}}_{(k)}(d)$ is the post-selected state
obtained by averaging over all accepted histories (defined explicitly
below).}

{Since fidelity is a linear functional of the state and
  $\rho^{\mathrm{mix}}_{(k)}(d)$ is a convex combination of
  conditional outputs, $F^{\mathrm{mix}}_{(k)}(d)$ reduces to the same
  convex combination of the corresponding conditional
  fidelities. Therefore, it suffices to specify three ingredients: (i)
  the conditional output state $\rho_{T,(k)}(\mathbf{y}_k)$ for a
  fixed history $\mathbf{y}_k = (y_1, \dots, y_k)$, (ii) its fidelity
  to the target SCS, and (iii) the statistical weight with which each
  history enters the mixture.}

{Following the same normalization convention as in
  Eq.~(\ref{eq:rhoAj}), the normalized conditional output at step $k$
  is obtained from the unnormalized matrix
  $\tilde{\rho}^{(k)}_T(\mathbf{y}_k)$ of Eq.~(\ref{eq:rhoj}) by
  dividing by the trace
  $\mathrm{Tr}_T[\tilde{\rho}^{(k)}_T(\mathbf{y}_k)] = P_k(y_k \mid
  \mathbf{y}_{<k})$:
\begin{equation}
    \rho_{T,(k)}(\mathbf{y}_k) = \frac{\tilde{\rho}^{(k)}_T(\mathbf{y}_k)}{P_k(y_k \mid \mathbf{y}_{<k})}.
    \label{eq:rho_T_k_normalized}
\end{equation}
Its fidelity with the target SCS, generalizing Eq.~(\ref{ab13}), is
the history-dependent matrix element
\begin{equation}
    F^{(k)}_{\mathrm{cat}}(\mathbf{y}_k)
    \equiv \langle \psi^{\mathrm{cat}(k)}_{\mathrm{out}} | \rho_{T,(k)}(\mathbf{y}_k) | \psi^{\mathrm{cat}(k)}_{\mathrm{out}} \rangle.
    \label{eq:F_cat_history}
\end{equation}
Each accepted history carries the normalized weight
$P(\mathbf{y}_k)/P_{(k)}(d)$, where the joint density
$P(\mathbf{y}_k)$ is related to the conditional densities by the chain
rule~(\ref{eq:chain_rule}). The post-selected mixed state is therefore
\begin{equation}
    \rho^{\mathrm{mix}}_{(k)}(d)
    = \frac{1}{P_{(k)}(d)} \int_{\mathcal{D}_d} d^k \mathbf{y}_k \, P(\mathbf{y}_k) \, \rho_{T,(k)}(\mathbf{y}_k),
    \label{eq:rho_mix}
\end{equation}
with $\mathcal{D}_d \equiv [-d/2, d/2]^k$ and $P_{(k)}(d)$ given by
Eq.~(\ref{R26}). Substituting~(\ref{eq:rho_mix})
into~(\ref{eq:F_mix_def}) and using linearity of the expectation value
to move the projector
$|\psi^{\mathrm{cat}(k)}_{\mathrm{out}}\rangle \langle
\psi^{\mathrm{cat}(k)}_{\mathrm{out}}|$ under the integral yields, in
view of~(\ref{eq:F_cat_history}), the weighted-average representation:
\begin{equation}
    F^{\mathrm{mix}}_{(k)}(d)
    = \frac{1}{P_{(k)}(d)} \int_{\mathcal{D}_d} d^k \mathbf{y}_k \, P(\mathbf{y}_k) \, F^{(k)}_{\mathrm{cat}}(\mathbf{y}_k).
    \label{eq:F_mix_weighted}
\end{equation}}

{Since at every iteration the target input is vacuum and the ancilla
  is the rotated pure conditional output of the preceding step, the
  normalized conditional output~(\ref{eq:rho_T_k_normalized}) is a
  rank-one operator
  $\rho_{T,(k)}(\mathbf{y}_k) =
  |\widetilde\psi_{(k)}(\mathbf{y}_k)\rangle
  \langle\widetilde\psi_{(k)}(\mathbf{y}_k)|
  \big/\|\widetilde\psi_{(k)}(\mathbf{y}_k)\|^{2}$, where
  $|\widetilde\psi_{(k)}(\mathbf{y}_k)\rangle$ is the unnormalized
  conditional output state [see Appendix~\ref{app1} (\ref{g6})], and
  therefore,
\begin{equation}
F_{\mathrm{cat}}^{(k)}(\mathbf y_k) = \frac{\big|\langle\psi^{\mathrm{cat}}_{(k)}|\widetilde\psi_{(k)}(\mathbf y_k)\rangle\big|^2}{\|\widetilde\psi_{(k)}(\mathbf{y}_k)\|^2}.
\label{g2}
\end{equation}}
{On the other hand, according to Appendix~\ref{app1} (\ref{eq:P_y_compact}), $P(\mathbf{y}_k) =\|\widetilde\psi_{(k)}(\mathbf{y}_k)\|^{2}/(2\,\mathcal{N}_{k})$, so that $P(\mathbf{y})$ and $F^{(k)}_{\mathrm{cat}}(\mathbf{y})$  contain the history-dependent norm $\|\widetilde\psi_{(k)}(\mathbf{y})\|^{2}$ in opposite positions. It cancels in their product, reducing the integrand
  of~(\ref{eq:F_mix_weighted}) to 
\begin{align}
\label{g3}
P(\mathbf y)\,F_{\mathrm{cat}}^{(k)}(\mathbf y)
=
\frac{\big|\langle\psi^{\mathrm{cat}}_{(k)}|\widetilde\psi_{(k)}(\mathbf y)\rangle\big|^2}{2\mathcal N_k}.
\end{align}
Thus, the exact mixed-state fidelity for arbitrary $k$ can be
evaluated from a $k$-dimensional integral whose integrand is the
squared overlap with the undistorted cat state, without any additional
history-dependent normalization factor.}

{As in the single-shot case, the expression is manifestly physical:
  $0\le F_{\mathrm{cat}}^{(k)}(\mathbf y)\le 1$ implies
  $0\le F^{\mathrm{mix}}_{(k)}(d)\le 1$.}



\begin{figure*}[t!]
\centering
  \begin{tabular}{c @{\qquad} c}
    \includegraphics[width=0.49\linewidth]{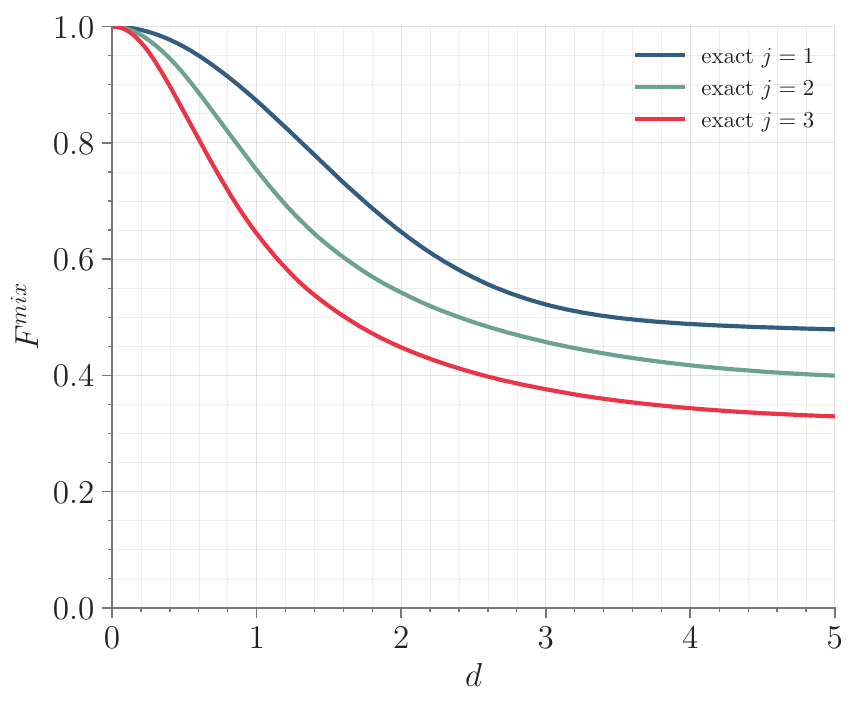} &
    \includegraphics[width=0.49\linewidth]{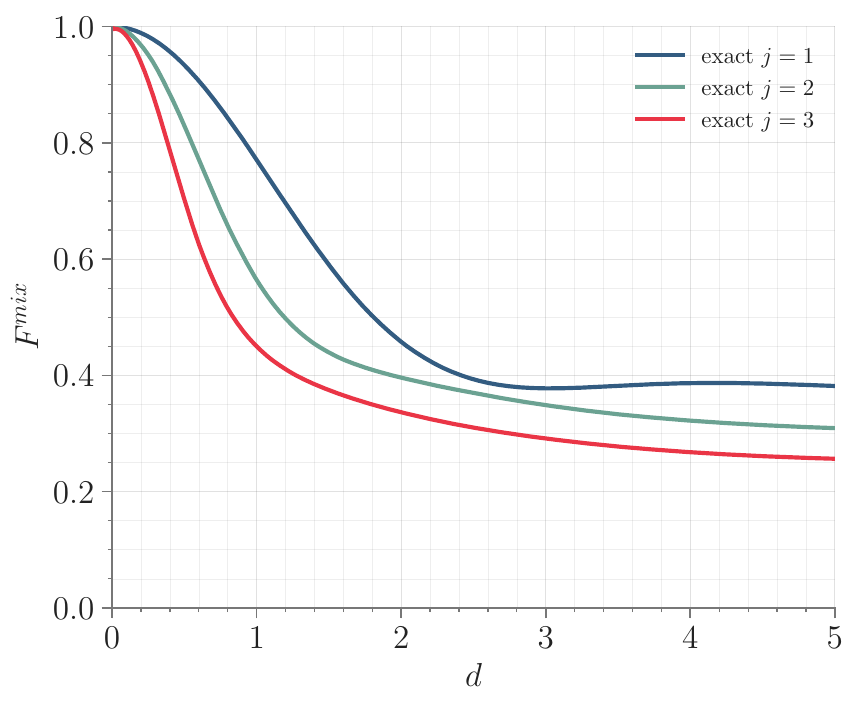}\\
    \small (a) Even initial SCS & \small (b) Odd initial SCS 
  \end{tabular}
  \caption{Post-selected mixed-state fidelity
    $F^{\mathrm{mix}}_{(k)}$ based on Eq.~(\ref{eq:F_mix_weighted})
    versus the acceptance-window width $d$ for $k=1,2,3$ number of
    iterations at fixed coupling $G=1.2$, with $a=\sqrt{2}$ and
    $r_0=1$ (vacuum target in each step). Panel (a) corresponds to an
    even initial SCS, panel (b) to an odd initial SCS.}
\label{fig:Fmix_vs_d}
\end{figure*}

\begin{figure*}[t!]
\centering
  \begin{tabular}{c @{\qquad} c @{\qquad} c}
    \includegraphics[width=0.32\linewidth]{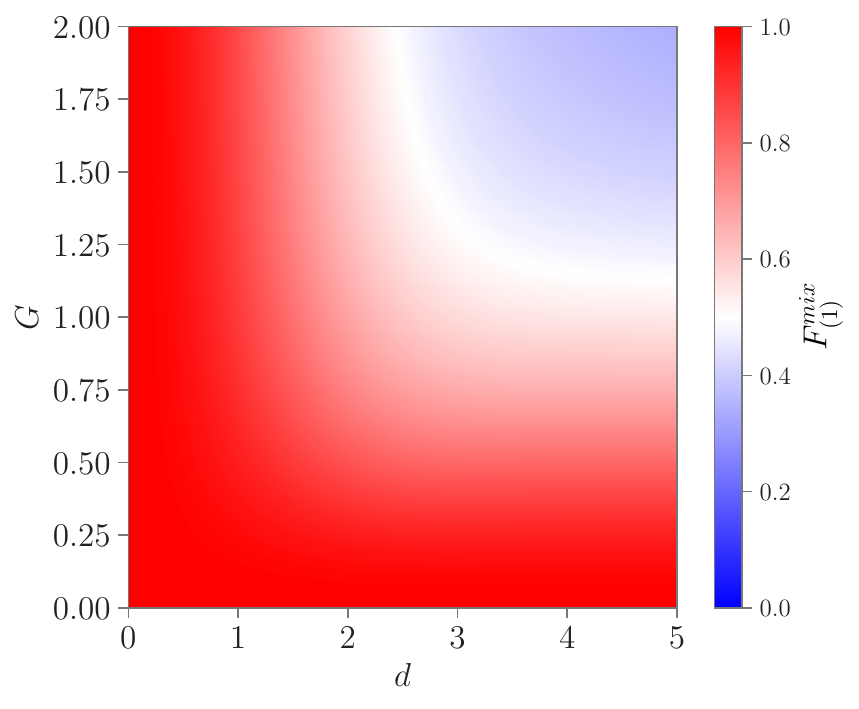} &
    \includegraphics[width=0.32\linewidth]{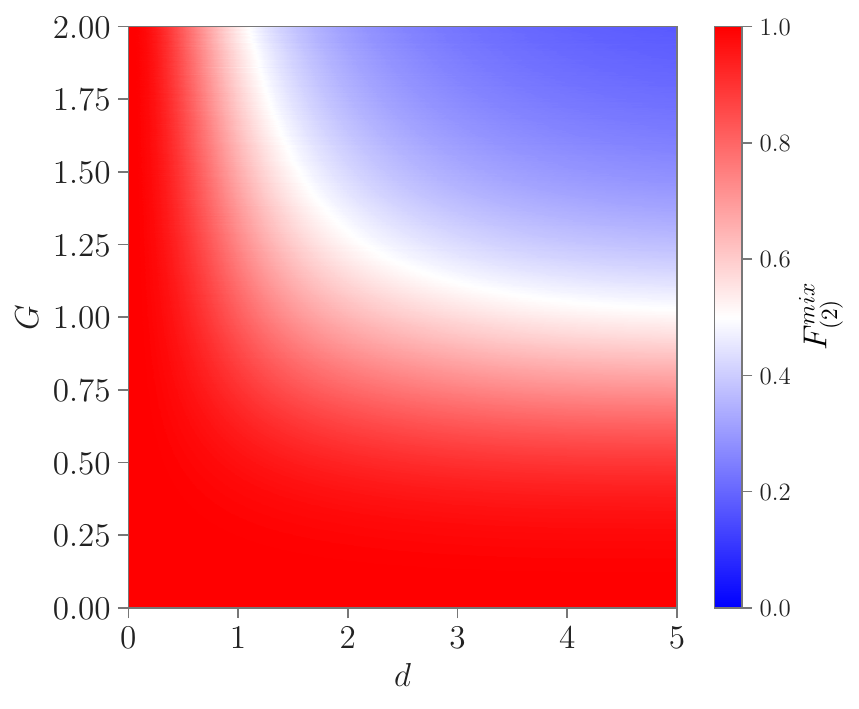} &
    \includegraphics[width=0.32\linewidth]{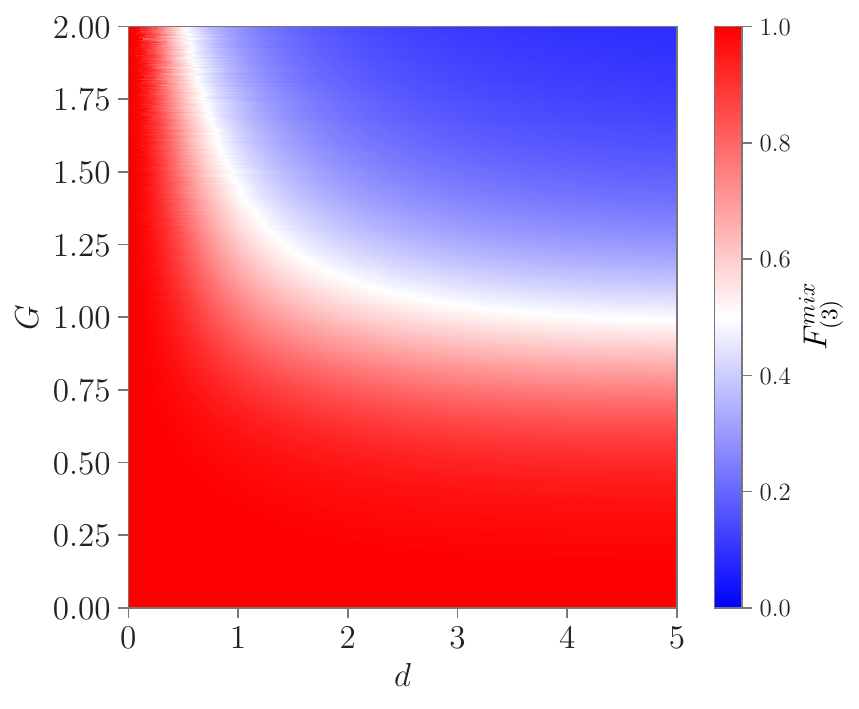} \\
    \small (a) $\,\,k=1$ & \small (b) $\,\,k=2$ & \small (c) $\,\,k=3$
  \end{tabular}
  \caption{Phase diagrams of the mixed-state fidelity
    $F^{\mathrm{mix}}_{(k)}(d,G)$ for an even initial ancillary SCS
    with $a=\sqrt{2}$ and $r_0=1$. Panels (a)--(c) correspond to $k=1$,
    $k=2$, and $k=3$ iterations, respectively. The color scale
    indicates $F^{\mathrm{mix}}_{(k)}$ ranging from $0$ (blue) to $1$
    (red).}
\label{fig:Fmix_phase_even}
\end{figure*}

\begin{figure*}[t!]
\centering
  \begin{tabular}{c @{\qquad} c @{\qquad} c}
    \includegraphics[width=0.32\linewidth]{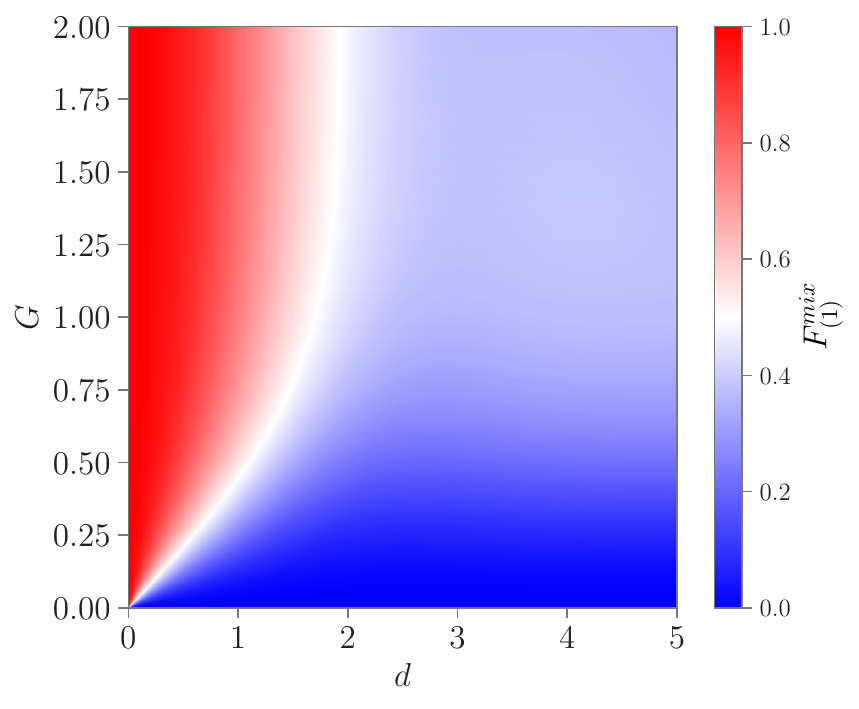} &
    \includegraphics[width=0.32\linewidth]{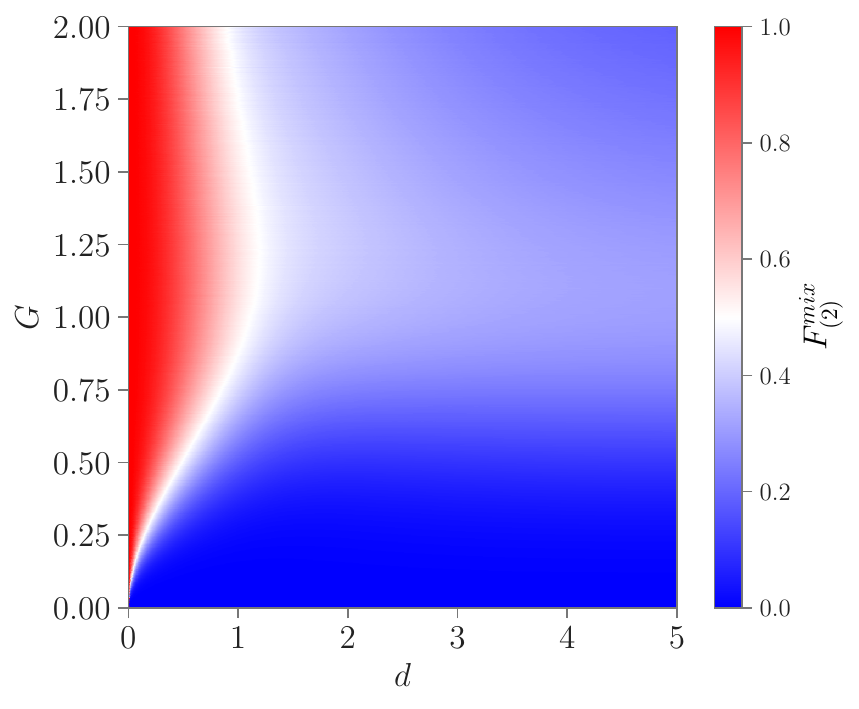} &
    \includegraphics[width=0.32\linewidth]{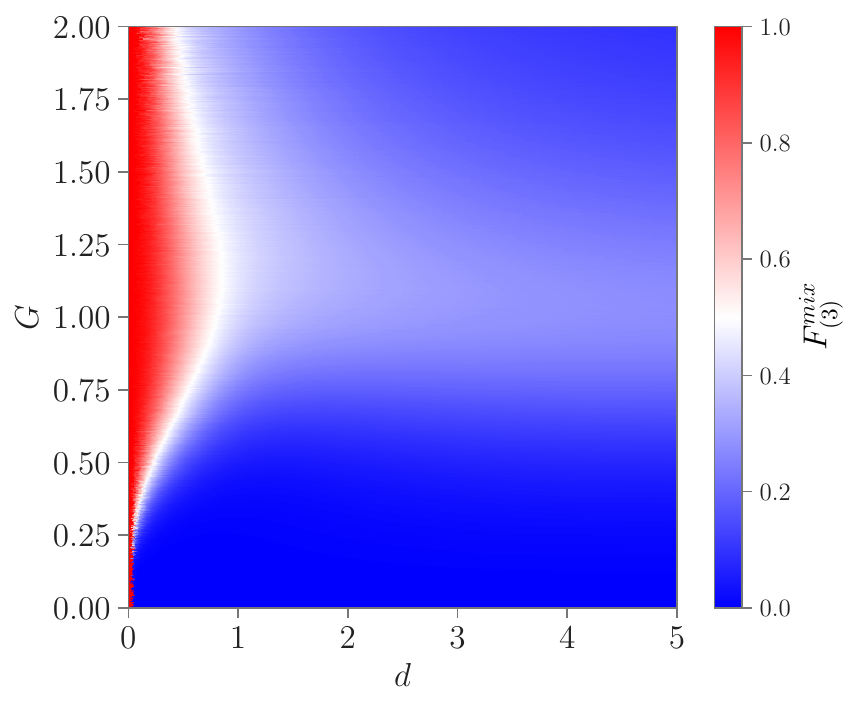} \\
    \small (a) $\,\,k=1$ & \small (b) $\,\,k=2$ & \small (c) $\,\,k=3$
  \end{tabular}
  \caption{Phase diagrams of the mixed-state fidelity
    $F^{\mathrm{mix}}_{(k)}(d,G)$ for an odd initial ancillary SCS
    with $a=\sqrt{2}$ and $r_0=1$. Panels (a)--(c) correspond to $k=1$,
    $k=2$, and $k=3$ iterations, respectively. The color scale
    indicates $F^{\mathrm{mix}}_{(k)}$ ranging from $0$ (blue) to $1$
    (red).}
\label{fig:Fmix_phase_odd}
\end{figure*}

{Figures~\ref{fig:Fmix_vs_d}--\ref{fig:Fmix_phase_odd} illustrate the
  dependence of the post-selected mixed-state
  fidelity~(\ref{eq:F_mix_weighted}) on the acceptance-window
  width~$d$. As shown in Fig.~\ref{fig:Fmix_vs_d} for $G=1.2$ and
  $k=1,2,3$, $F^{\mathrm{mix}}_{(k)}$ approaches unity as $d\to 0$ and
  generally decreases with~$d$, the degradation accelerating with~$k$
  due to the accumulation of history-dependent distortions. For
  every~$k$, the even initial SCS~[panel~(a)] sustains a noticeably
  higher fidelity at large~$d$ than the odd one~[panel~(b)], which can
  be traced to the parity-dependent structure of the probability
  density (cf.\ Fig.~\ref{figL2}): the even-parity density peaks at
  the ideal outcome $y_m = Gx_0 = 0$, whereas the odd-parity density
  exhibits a local minimum there, causing accepted odd-parity
  histories to depart from the ideal condition more rapidly as
  $d$~grows.  The phase diagrams $F^{\mathrm{mix}}_{(k)}(d,G)$ in
  Figs.~\ref{fig:Fmix_phase_even} and~\ref{fig:Fmix_phase_odd} confirm
  this parity asymmetry: the high-fidelity region of the even SCS
  spans a broad range of~$d$ and recedes only gradually with~$k$,
  while for the odd SCS it is confined to a narrow strip near $d = 0$
  already at $k = 1$ and contracts rapidly thereafter. In both cases
  the high-fidelity domain shrinks toward smaller~$d$ with increasing
  $|G|$, reflecting the heightened sensitivity of the conditional
  output to deviations from the ideal measurement outcome at stronger
  QND coupling.}

{This parity-dependent behavior of $F^{\mathrm{mix}}_{(k)}$ can be understood
directly from the structure of Eq.~(\ref{eq:F_mix_weighted}), which
expresses $F^{\mathrm{mix}}_{(k)}$ as the average of the conditional fidelity
$F^{(k)}_{\mathrm{cat}}(\mathbf{y}_k)$ weighted by the joint probability
density $P(\mathbf{y}_k)$. Since $F^{(k)}_{\mathrm{cat}}$ attains its maximum
at the ideal outcome $\mathbf{y}_k = \mathbf{0}$ and decreases with
increasing $|\mathbf{y}_k|$, the mixed-state fidelity is governed by how
strongly $P(\mathbf{y}_k)$ is concentrated near $\mathbf{y}_k = \mathbf{0}$.
For the even ancilla, $P$ peaks at the ideal outcome
(cf.\ Fig.~\ref{figL2}a), so the dominant contribution to the
weighted average comes from histories with near-unit fidelity;
consequently, $F^{\mathrm{mix}}_{(k)}$ remains high over an appreciable range
of~$d$. For the odd ancilla, $P$ exhibits a local minimum at
$\mathbf{y}_k = \mathbf{0}$ (cf.\ Fig.\ref{figL2}b), shifting the
dominant statistical weight toward non-ideal outcomes where
$F^{(k)}_{\mathrm{cat}}$ is already reduced, which results in a markedly
faster degradation of $F^{\mathrm{mix}}_{(k)}$ with~$d$.}


In the limit $d\to\infty$, the acceptance domain $\mathcal{D}_d$
expands to $\mathbb{R}^k$ and $P_{(k)}(d)\to 1$, so that
$F^{\rm mix}_{(k)}(d)$ saturates at the finite constant
$F^{\rm mix}_{(k)}(\infty) = \int_{\mathbb{R}^k}
P(\mathbf{y}_k)\,F^{(k)}_{\rm
  cat}(\mathbf{y}_k)\,\mathrm{d}^k\mathbf{y}_k$, which accounts for
the plateaus visible in Fig.~\ref{fig:Fmix_vs_d}. Furthermore, strict
monotonicity of $F^{\rm mix}_{(k)}(d)$ in $d$ is not
guaranteed. Differentiating Eq.~(\ref{eq:F_mix_weighted}) with respect
to $d$ reveals that $\mathrm{d}F^{\rm mix}_{(k)}/\mathrm{d}d$ changes
sign depending on whether the measurement histories newly admitted by
widening the acceptance window carry a conditional fidelity
$F^{(k)}_{\rm cat}$ above or below the current weighted average
$F^{\rm mix}_{(k)}(d)$. Since $F^{(k)}_{\rm cat}(\mathbf{y}_k)$
involves the interference phase $\Psi_k(\mathbf{y}_k)$
[Eq.~(\ref{z888})], which depends linearly on the measurement
outcomes, the conditional fidelity inherits an oscillatory dependence
on $\mathbf{y}_k$ whose local maxima at nonzero outcomes can exceed
the running average, giving rise to weak non-monotonic features. At
$k=1$, this oscillatory structure reduces to the explicit $\cos(ay_m)$
and $\sin(ay_m)$ factors of Eqs.~(\ref{ab15})--(\ref{ab18}) and is
most pronounced, although the Gaussian envelope
$\exp(-\mu y_m^2/2r^4)$ confines such deviations to a narrow range
of~$d$. For $k>1$, the higher-dimensional averaging progressively
smooths them out.

{The behavior of $F^{\mathrm{mix}}_{(k)}$ and $P_{(k)}$ as functions of
the acceptance-window width~$d$ reveals a fundamental trade-off that
governs the operating regime of the protocol. For $|G|\gtrsim 1$, the
two figures of merit follow opposite overall trends in~$d$: widening
the window raises $P_{(k)}$ by admitting more measurement outcomes, but
simultaneously lowers $F^{\mathrm{mix}}_{(k)}$ by including histories that
deviate further from the ideal condition $y_m = Gx_0 = 0$. An operating
point that secures simultaneously high $F^{\mathrm{mix}}_{(k)}$ and
appreciable $P_{(k)}$ is therefore confined to a limited interval of~$d$
whose position and width are set jointly by the ancillary size~$a$,
the coupling~$G$, and the iteration order~$k$. With increasing~$k$,
this interval contracts and the figures of merit attainable within it
degrade, which reflects the cumulative impact of finite homodyne
resolution on the iterated protocol. The contraction is markedly more
pronounced for odd-parity than for even-parity initial ancillas, so
that, from the viewpoint of practical implementation, an even initial
ancilla is preferable: it tolerates a broader acceptance window and
therefore offers more flexibility in balancing fidelity against success
probability. A complementary strategy to enlarge the usable
regime---especially at large~$k$---is to decrease $|G|$, which slows
the per-step growth of the cat size $|G|^k a$ and of the accumulated
squeezing factor $r_{(k)} = \sqrt{\sum_{l=0}^{k} G^{2l}}$, but relaxes
the sensitivity of both $F^{\mathrm{mix}}_{(k)}$ and $P_{(k)}$ to the
homodyne resolution. Together, these observations define a practical
route to optimizing the number of iterations, the coupling strength, and
the postselection window against a given experimental resolution and
target cat-state parameters.}

\section{Discussions and conclusion}
\label{conclusion}

We have investigated the CV homodyne-conditioned $C_Z$-gate–based
protocol capable of generating a two-component squeezed
Schr{\"o}dinger cat state–like superposition from vacuum and a
small-amplitude cat state at the input, as well as an iterative
cat-state amplification scheme built upon this protocol. The scheme
relies on conditional measurements on ancillary modes, which provide
the effective nonlinearity required for the preparation of
non-Gaussian states in continuous-variable optical systems. We have
identified the operational regime of the gate in which the output
state is an undistorted squeezed Schr{\"o}dinger cat state. Both the
squeezing and the effective size of the generated Schr{\"o}dinger
cat state at the gate output, quantified by the separation between
the coherent components, can be increased in a controlled manner
through an appropriate choice of the parameters: QND coupling strength
$G$ and the target and ancillary input states squeezing parameters.

Iterative measurement-induced cat-state engineering can be naturally
integrated with standard Gaussian operations—including displacement,
phase-space rotation, squeezing, and shear transformations—forming a
versatile hybrid framework for non-Gaussian state preparation. Such
hybrid approach provides a scalable route for generating squeezed
Schr{\"o}dinger cat states and incorporating them as resources in
larger non-Gaussian continuous-variable quantum networks and photonic
quantum information architectures.

\begin{acknowledgments}
The work was supported by Russian
Science Foundation (project No. 24-11-00398).
\end{acknowledgments}

\section*{Declaration of competing interests}

The authors declare that they have no known competing financial
interests or personal relationships that could have appeared to
influence the work reported in this paper.

\textbf{Data Availability Statement}: This theoretical study has no associated experimental or numerical dataset.

%

\appendix
\section{Joint probability density after iterations}
\label{app1}

{In this Appendix we derive a closed-form expression for the joint
  probability density $P(y_1,...,y_k)$ of all homodyne measurement
  outcomes in the $k$-step iterative cat-state growth protocol.}

{Throughout, we denote position-representation wave functions as
  $\psi(x)\equiv\langle x|\psi\rangle$ and momentum-representation
  wave functions as ${\tilde\psi}(p)\equiv\langle p|\psi\rangle$. The
  two are related by the Fourier transform
  ${\tilde\psi}(p)=(2\pi)^{-1/2}\int e^{-ipx}\psi(x)\,dx$. We set
	  $\hbar=1$, so $[\hat x,\hat p]=i$ and $a\equiv\sqrt{2}\,\alpha$ is
	  \textcolor{black}{the cat-state size parameter associated with} the coherent state
  $|\alpha\rangle$.}

{The initial ancillary cat state
  $|\mathrm{cat}\rangle=\mathcal{N}(|{+}\alpha\rangle+s|{-}\alpha\rangle)$,
  with $s=\pm1$ for the even/odd superposition, has the density matrix
\begin{gather}
\rho_{\mathrm{cat}}=\mathcal{N}^2\sum_{\sigma,\sigma'=\pm1}c_{\sigma\sigma'}|\sigma\alpha\rangle\langle\sigma'\alpha|,
\label{Aa1}
\end{gather}
where the coefficients are $c_{\pm\pm}=1$, $c_{\pm\mp}=s$, and the sum
runs over the four combinations $(\sigma,\sigma')$. Each term
$|\sigma\alpha\rangle\langle\sigma'\alpha|$ corresponds to one element
of the $2\times2$ coherent-state block structure of
$\rho_{\mathrm{cat}}$; we refer to a branch as the single
coherent-state component labeled by one value of $\sigma$ in the ket
side, i.e., the state $|\sigma\alpha\rangle$ with $\sigma=+1$ or
$\sigma=-1$. Thus, the density matrix contains four matrix elements
built from two branches. Since all subsequent operations — the QND
gate $\hat C_Z=e^{iG\hat q_T\hat q_A}$, homodyne projection, and a
$-\pi/2$ rotation — are linear in the density matrix, each branch
$\sigma$ propagates independently through the $k$-step protocol,
yielding an unnormalized wave function $h_\sigma^{(k)}(x)$ in the
position representation of the surviving (target) mode. The joint
probability density of all homodyne outcomes
$\mathbf y_k\equiv(y_1,\dots,y_k)$ is then assembled from the overlaps
of these branch wave functions:
\begin{gather}
P(\mathbf y_k)=\frac{1}{2\mathcal{N}_k}\sum_{\sigma,\sigma'=\pm}c_{\sigma\sigma'}\,\langle h_{\sigma'}^{(k)}|h_\sigma^{(k)}\rangle,
\label{Aa2}
\end{gather}
where $\mathcal{N}_k$ is a single global normalization constant.}

{At each iteration $j$, the target input is vacuum,
  $\psi_0(x)=\pi^{-1/4}e^{-x^2/2}$, and the ancilla is the
  $-\pi/2$-rotated output of iteration $j{-}1$. The QND gate acts in
  the joint position representation as
  $\langle x_T|\otimes\langle x_A|\hat
  C_Z|\Psi\rangle=e^{iGx_Tx_A}\Psi(x_T,x_A)$. Projecting the ancilla
  onto the momentum eigenstate $|y_j\rangle$ requires computing the
  Fourier transform over $x_A$ via the relation
  $\langle y_j|x_A\rangle=(2\pi)^{-1/2}e^{-iy_jx_A},$ which converts
  the integral over $x_A$ into the ancilla's momentum-representation
  wave function evaluated at the shifted argument $y_j-Gx_T$. Denoting
  the ancilla's momentum-representation wave function at step $j-1$ by
  ${\tilde\psi}_A^{(j-1,\sigma)}(p)$ (i.e., the rotated
  branch-$\sigma$ output of step $j{-}1$), we obtain
\begin{gather} 
h^{(j)}_\sigma(x_T)=\pi^{-1/4}\,e^{-x_T^2/2}\;{\tilde\psi}_A^{(j-1,\sigma)}\!\left(y_j-Gx_T\right).
\label{S1}
\end{gather}}

{The phase-space rotation operator $\hat R=e^{i\pi\hat n/2}$ acts
  according to $\hat R^\dagger\hat{x}\hat R=-\hat{p}$ and
  $\hat R^\dagger\hat{p}\hat R=\hat{x}$, and therefore transforms
  position eigenstates into momentum eigenstates:
  $\hat R|x\rangle=|p{=}x\rangle$. Consequently, for any state
  $|\psi\rangle$ with position wave function $\psi(x)$, the rotated
  state $\hat R|\psi\rangle$ has the momentum-representation wave
  function
\begin{gather} 
\langle p|\hat R|\psi\rangle = \int dx\,\psi(x)\langle p|p{=}x\rangle = \psi(p).
\label{S2}
\end{gather}
In other words, the momentum-representation wave function of the
rotated state equals numerically the position-representation wave
function $\psi$ of the original (unrotated) state evaluated at the
same argument $p$. Substituting this result into Eq.~(\ref{S1}) and
expressing the ancilla's momentum wave function through the position
wave function of the previous step's output, $h^{(j-1)}_\sigma(x)$, we
arrive at the closed recursion
\begin{gather}
h^{(j)}_\sigma(x)=\pi^{-1/4}\,e^{-x^2/2}\;h^{(j-1)}_\sigma(y_j-Gx),\; j=1,2,\dots,k.
\label{S3}
\end{gather}
This relation makes the computational structure of the protocol
explicit: no explicit Fourier transform is needed at any intermediate
step because the $-\pi/2$ rotation converts the
position-representation wave function of the output into the momentum
wave function of the next step ancilla.}

{In the first step ($j=1$) the ancilla is the initial cat state
  itself. The branch-$\sigma$ component is the coherent state
  $|\sigma\alpha\rangle$, whose momentum-representation wave function
  reads
\begin{gather}
{\tilde\psi}_{\sigma\alpha}(p)=\pi^{-1/4}\exp\!\left(-\frac{p^2}{2}-i\sigma a\,p\right),\qquad a=\sqrt{2}\alpha.
\label{S4}
\end{gather}
We define $h^{(0)}_\sigma(p)\equiv{\tilde\psi}_{\sigma\alpha}(p)$ as the initial condition for the recursion~(\ref{S3}). For all $j\geq1$, $h^{(j)}_\sigma(x)$ is the position-representation wave function of the target mode surviving after step $j$.}

{To solve the recursion~(\ref{S3}) we introduce the auxiliary linear
  functions:
\begin{gather}
\zeta_0(x)=x,\qquad \zeta_l(x)=y_{k+1-l}-G\,\zeta_{l-1}(x),\quad l=1,\dots,k,
\label{S5}
\end{gather}
so that $\zeta_l(x)=(-G)^l x+\phi_l$ with shifts $\phi_0=0$,
$\phi_l=y_{k+1-l}-G\phi_{l-1}$. These functions encode the nested
structure of the $k$ successive measurements. The unnormalized
branch wave function after $k$ iterations takes the form (in the
position representation of the surviving mode):
\begin{gather}
h^{(k)}_\sigma(x)=\pi^{-(k+1)/4}\exp\!\left(-\frac{1}{2}\sum_{l=0}^k\zeta_l^2(x)\right)\exp\!\left(-i\sigma a\,\zeta_k(x)\right).
\label{S6}
\end{gather}
The proof proceeds by induction. For the base case $k=1$, we have
$\zeta_0=x$ and $\zeta_1=y_1-Gx$, and direct calculation using
Eq.~(\ref{S3}) with the initial condition~(\ref{S4}) yields
\begin{gather}
h_\sigma^{(1)}(x)=\pi^{-1/2}e^{-(\zeta_0^2+\zeta_1^2)/2}e^{-i\sigma a\zeta_1},
\label{S60}
\end{gather}
in agreement with Eq.~(\ref{S6}). For the inductive step, assume the
formula holds at step $k-1$ with outcomes $y_1, \ldots, y_{k-1}$. The
auxiliary functions for the $(k-1)$-step protocol, which we denote by
$\zeta^{(k-1)}_l$ to distinguish them from
$\zeta_l \equiv \zeta^{(k)}_l$, are built from the same recursion
(\ref{S5}) but with $k$ replaced by $k-1$:
{\color{black}
\begin{align}
\zeta^{(k-1)}_0(x')&=x', \notag\\
 \zeta^{(k-1)}_l(x')&=y_{k-l}-G\,\zeta^{(k-1)}_{l-1}(x'),\quad l=1,\dots,k-1, 
\label{S50}
\end{align}}
Substituting the induction hypothesis into the recursion (\ref{S3}) at
step $k$ and setting $x' = y_k - Gx$, one finds that the two sets of
auxiliary functions are linked by the identity
$\zeta^{(k-1)}_l(y_k - Gx) = \zeta^{(k)}_{l+1}(x)$
$(l = 0,1,\ldots,k-1)$. Using this identity together with
$x = \zeta_{0}(x)$, the sum in the Gaussian exponent becomes
$\sum_{l=0}^k \zeta_l^2(x)$ and the phase becomes
$-i\sigma a\,\zeta_k(x)$, reproducing Eq.~(\ref{S6}) at step $k$.}

{For notational brevity, the dependence on the measurement history is
  suppressed: the branch wave function $h^{(k)}_{\sigma}(x)$ is, in
  fact, a function of $x$ and of the full outcome vector
  $\mathbf{y}_{k}=(y_{1},\dots,y_{k})$, entering through the auxiliary
  functions $\zeta_{l}(x)$ defined in Eq.~(\ref{S5}). We shall write
  $h^{(k)}_{\sigma}(x,\mathbf{y}_{k})$ whenever this dependence needs
  to be made explicit.}

{We next obtain expressions for overlaps and the joint probability
  density. Since $|h^{(k)}_\sigma(x)|^2$ is independent of $\sigma$
  (the $\pm\sigma$ phases cancel in the modulus squared), the diagonal
  overlaps are equal:
\begin{gather}
W_k\equiv\|h^{(k)}_+\|^2=\|h^{(k)}_-\|^2=\frac{1}{\pi^{k/2}\sqrt{A_k}}\exp\!\left(\frac{E_k^2}{A_k}-F_k\right),
\label{S7}
\end{gather}
where $\|h^{(k)}_\pm\|^2\equiv\langle h^{(k)}_\pm|h^{(k)}_\pm\rangle$, $A_k=\sum_{l=0}^k G^{2l}$, $E_k=\sum_{l=1}^k(-G)^l\phi_l$, and $F_k=\sum_{l=1}^k\phi_l^2$. Physically, $W_k$ is the outcome-dependent Gaussian envelope that determines the overall probability scale and is identical for both branches.}

{The off-diagonal overlap, which encodes the quantum interference
  between the two coherent-state components, is computed by Gaussian
  integration with an imaginary linear shift:
\begin{gather}
\langle h^{(k)}_-|h^{(k)}_+\rangle = W_k\,\mathcal{V}_k\,e^{-i\Psi_k},
\label{S8}
\end{gather}
  \textcolor{black}{with the outcome-independent interference visibility}
\begin{gather}
\mathcal{V}_k=\exp\!\left(-\frac{a^2 G^{2k}}{A_k}\right),
\label{S9}
\end{gather}
and the interference phase (linear in the outcomes):
\begin{gather}
\Psi_k(y_1,\dots,y_k)=2a\!\left[\phi_k-\frac{(-G)^k E_k}{A_k}\right].
\label{S10}
\end{gather}
\textcolor{black}{The factor $\mathcal{V}_k$ is controlled by the ratio between the squared
component displacement $G^{2k}a^2$ and the accumulated squeezing parameter
$A_k$. It should not be interpreted as physical decoherence: after
integration over all outcomes it combines with the phase averaging to
preserve the fixed initial branch overlap $e^{-a^2}$.}}

{Collecting all four terms of the branch decomposition (\ref{Aa2}), we obtain the joint probability density:
\begin{align}
  \label{S11}
  P(\mathbf y_k)=\frac{1}{\mathcal{N}_k}\,W_k(\mathbf y_k) \left[1+s\,\mathcal{V}_k\cos\Psi_k(\mathbf y_k)\right],
\end{align}
where $\mathcal{N}_k=\int_{\mathbb{R}^k}\!W_k[1+s\,\mathcal{V}_k\cos\Psi_k]\,d^ky$ is the global normalization constant. As shown in Appendix~\ref{app2}, $\mathcal{N}_k$ is independent of $k$ and equals $1+s e^{-a^2}$ for all $k$.}

{The structure of Eq.~(\ref{S11}) admits a transparent physical
  interpretation. The Gaussian factor $W_k$ describes the classically
  expected distribution of measurement outcomes, while the oscillatory
  factor $1+s\,\mathcal{V}_k\cos\Psi_k(\mathbf y_k)$ encodes the
  non-classical interference between the two coherent-state branches
  \textcolor{black}{of the cat state. The outcome-independent factor
  $\mathcal{V}_k$ controls the visibility of this conditional
  interference pattern as the branches evolve in phase space, while
  the full outcome-averaged branch overlap remains fixed as discussed
  above.}}

{To achieve a more compact form of the joint probability density, it
  is convenient to introduce the unnormalized conditional output state
\begin{equation}
|\widetilde\psi_{(k)}(\mathbf{y}_k)\rangle
= |h^{(k)}_{+}(\mathbf{y}_k)\rangle + s\,|h^{(k)}_{-}(\mathbf{y}_k)\rangle,
\label{g6}
\end{equation}
built from the two branch position-representation wave functions
$h^{(k)}_{\sigma}(x,\mathbf{y}_k)$ [see Eq.~(\ref{S6})]; it is the
$k$-step analogue of $|\widetilde\psi_{\mathrm{out}}\rangle$ of
Eq.~(\ref{a7}). The branch identity
$\|\widetilde\psi_{(k)}(\mathbf{y}_k)\|^{2} =
2\,W_{k}(\mathbf{y}_k)\,[1 +
s\,\mathcal{V}_k\cos\Psi_{k}(\mathbf{y}_k)]$ turns the joint
density~(\ref{S11}) into the form
\begin{equation}
P(\mathbf{y}_k) = \frac{\|\widetilde\psi_{(k)}(\mathbf{y}_k)\|^{2}}{2\,\mathcal{N}_{k}},
\label{eq:P_y_compact}
\end{equation}}

{Consider the case of a single iteration $k=1$. We have $\phi_1=y_1$,
  $E_1=-Gy_1$, giving $\Psi_1=2ay_1/r_{(1)}^2$ with
  $r_{(1)}^2=1+G^2$. The joint density reduces to
\begin{align}
  \label{S101}
  P(y_1)\!=2\mathcal{N}^2 W_1\!\left[1+s\,e^{-a^2 G^2/r_{(1)}^2}\cos\!\left(\frac{2a\,y_1}{r_{(1)}^2}\right)\!\right],
\end{align}
where $W_1=e^{-y_1^2/r_{(1)}^2}\textcolor{black}{/(\sqrt{\pi}\, r_{(1)})},\,\, {\mathcal
  N}=[2(1+se^{-a^2})]^{-1/2}$ is
the normalization factor of the cat-state, which corresponds to Eq.~(\ref{a0017}) of the main text (for $x_0=0$, $r=\delta=1$).}

\section{Total Success Probability}
\label{app2}

{In this Appendix we evaluate the total probability $P_{(k)}(d)$ that
  all $k$ homodyne outcomes fall within the acceptance window
  $[-d/2,d/2]$, starting from the joint probability density derived in
  Appendix~\ref{app1},
\begin{gather}
P_{(k)}(d) = \int_{|y_1|\le d/2}\!\cdots\!\int_{|y_k|\le d/2} P(y_1,\dots,y_k)\,dy_1\cdots dy_k,
\label{z1}
\end{gather}
The key idea is to re-express both the Gaussian envelope $W_k$ and the
interference phase $\Psi_k$ in a form that reflects the sequential,
measurement-by-measurement structure of the protocol. \textcolor{black}{This recasts the
$k$-dimensional integral as a nested sequence of conditional
one-variable quadratures} and reveals that the global normalization
is independent of the iteration number.}
 
{The multivariate Gaussian envelope $W_k(\mathbf y_k)$ derived in
  Appendix~\ref{app1} [Eq.~(\ref{S7})] depends on all $k$ outcomes
  simultaneously through the quadratic form $F_k- E_k^2/A_k$. However,
  since the protocol is sequential—each measurement conditions the
  state for the next one—it is natural to decompose $W_k$ into a
  product of conditional single-variable Gaussians (the Cholesky
  decomposition):
\begin{gather}
W_k(\mathbf y_k) = \prod_{j=1}^{k}\,\frac{\eta_j}{\sqrt\pi}\,\exp\!\Big[-\eta_j^2\big(y_j-\mu_j\big)^2\Big], 
\label{z3}
\end{gather}
with the conditional inverse-width parameters
\begin{gather}
\eta_j = \frac{r_{(j-1)}}{r_{(j)}},\qquad r_{(j)}^2=\sum_{l=0}^{j}G^{2l},\quad r_{(0)}=1,
\label{z4}
\end{gather}
and the conditional means defined recursively:
\begin{gather}
\mu_1=0,\qquad \mu_{j+1}=\frac{G\,r_{(j-1)}^2}{r_{(j)}^2}\,(y_j-\mu_j),\quad j=1,\dots,k{-}1.
\label{z5}
\end{gather}
Here $r_{(j)}$ is the inverse quadrature squeezing factor after $j$
iterations (Sec.~\ref{section_III} of the main text). The parameter
$\eta_j=r_{(j-1)}/r_{(j)}$ is the inverse-width parameter of the
conditional Gaussian distribution for the $j$-th homodyne outcome.
Since $r_{(j)}^2-r_{(j-1)}^2=G^{2j}$, one has $\eta_j<1$ for
$G\neq0$. Hence, the standard deviation of the conditional outcome
distribution is larger than the vacuum homodyne standard deviation by
the factor $1/\eta_j=r_{(j)}/r_{(j-1)}$. 
The recursion (\ref{z5}) encodes how each measurement outcome shifts the
conditional center of the next distribution: the deviation
$(y_j-\mu_j)$ is propagated through the QND coupling $G$ and
attenuated by $r^2_{(j-1)}/r^2_{(j)}$, because the additional
squeezing introduced at step $j+1$ absorbs part of the displacement
information. The initial condition $\mu_1 =0$ follows from the
symmetry of the vacuum target and the initial cat ancilla.  The
correctness of Eq.~(\ref{z3}) is verified by checking that the
telescoping product
$\prod_{j=1}^k \eta_j = r_{(0)}/r_{(k)} = 1/\sqrt{A_k}$ reproduces the
prefactor of $W_k$, and that $\sum_{j=1}^k\eta_j^2(y_j{-}\mu_j)^2$
reconstructs the exponent $F_k- E_k^2/A_k$.}

{The interference phase $\Psi_k$ [Eq.~(\ref{S10})], which encodes the
  quantum coherence between the two cat-state branches, decomposes
  additively in the same conditional (Cholesky) variables:
\begin{gather}
\Psi_k(y_1,\dots,y_k) = \sum_{j=1}^{k}\delta_j\,(y_j-\mu_j),\qquad \delta_j\equiv\frac{2a\,(-G)^{j-1}}{r_{(j)}^2}. 
\label{z6}
\end{gather}
Each coefficient $\delta_j$ is the sensitivity of the interference
phase to the $j$-th measurement deviation. The factor $(-G)^{j-1}$
reflects the alternating sign of the QND coupling accumulated over
$j-1$ intermediate $-\pi/2$ rotations, while the denominator
$r^2_{(j)}$ accounts for the dilution of sensitivity by the growing
squeezing. As a result, $|\delta_j|$ decreases with $j$: the
interference pattern becomes progressively less sensitive to later
measurement outcomes as the cat-state components separate further in
phase space.}

{Combined with Eq.~(\ref{z3}), the additive structure ensures that the
  integrand $W_k e^{i\Psi_k}$ factorizes into a product of
  conditional one-variable factors, which is the basis for the
  sequential evaluation below.}

{Substituting the joint density~(\ref{S11}) into the definition of
  $P_{(k)}(d)$ and separating the classical (diagonal) and quantum
  (off-diagonal) contributions from the branch
  decomposition~(\ref{Aa2}), we obtain
\begin{gather}
P_{(k)}(d) = \frac{1}{\mathcal{N}_k}\left[J_k^{(0)}(d)+s\,\mathcal{V}_k\;\mathrm{Re}\,J_k^{(1)}(d)\right],
\label{z7}
\end{gather}
where the Gaussian integral and the interference integral are defined as
\begin{gather}
J_k^{(0)}(d) = \int_{\mathcal{D}_d} W_k\,d^k\!y,\qquad J_k^{(1)}(d) = \int_{\mathcal{D}_d} W_k\,e^{i\Psi_k}\,d^k\!y, 
\label{z8}
\end{gather}
with $\mathcal{D}_d\equiv[-d/2,d/2]^k$. The Gaussian integral
$J_k^{(0)}$ accounts for the classical mixture of the two
coherent-state components (diagonal density-matrix elements), while
the interference integral $J_k^{(1)}$, weighted by the visibility
$\mathcal{V}_k$ and taken as a real part, captures the quantum
coherence between the branches (off-diagonal elements).}
 
{Setting $d\to\infty$ and requiring $P_{(k)}(\infty)=1$, we find the
  global normalization constant $\mathcal{N}_k$ from the full-space
  integrals. Since $W_k$ is normalized, $J_k^{(0)}(\infty)=1$. For
  $J_k^{(1)}(\infty)$, the factorized form of $W_k\,e^{i\Psi_k}$
  yields a product of standard Gaussian integrals with imaginary
  linear shifts:
\begin{gather}
J_k^{(1)}(\infty) = \prod_{j=1}^{k}e^{-\delta_j^2/(4\eta_j^2)},\qquad \frac{\delta_j^2}{4\eta_j^2} = \frac{a^2 G^{2(j-1)}}{r_{(j-1)}^2\,r_{(j)}^2}. 
\label{z9}
\end{gather}
\textcolor{black}{Using $r_{(j)}^2-G^2r_{(j-1)}^2=1$, each term can be written as
$a^2[G^{2(j-1)}/r_{(j-1)}^2-G^{2j}/r_{(j)}^2]$. Hence}
$\sum_{j=1}^k \delta_j^2/(4\eta_j^2)$ telescopes to
\begin{gather}
\sum_{j=1}^k \frac{\delta_j^2}{4\eta_j^2}
= a^2\,\frac{A_k-G^{2k}}{A_k}
= a^2\left(1-\frac{G^{2k}}{A_k}\right).
\label{z9a}
\end{gather}
Combined with the visibility $\mathcal{V}_k = e^{-a^2 G^{2k}/A_k}$, this gives
\begin{gather}
\mathcal{V}_k\,J_k^{(1)}(\infty) = e^{-a^2}=e^{-2\alpha^2}, 
\label{z10}
\end{gather}
and therefore
\begin{gather}
\mathcal{N}_k = 1+s\,e^{-a^2},
\label{z11}
\end{gather}
for all $k$.}

{The independence of the normalization constant $\mathcal{N}_k$ from
  $k$ has a transparent physical origin. The total probability, summed
  over all possible measurement outcomes at every step, must equal
  unity. Since neither the QND entangling operations nor the $-\pi/2$
  rotations alter the trace of the quantum state, and since each
  homodyne projection provides a complete resolution of the identity,
  the only quantity that enters the normalization is the overlap
  $\langle{+}\alpha|{-}\alpha\rangle=e^{-2\alpha^2}$ between the two
  coherent-state branches of the initial cat state. This overlap is
  fixed at the beginning of the protocol and does not change with the
  number of iterations. In particular,
  $\mathcal{N}_k=\mathcal{N}^{-2}_{\pm}/2$, where
  $\mathcal{N}_{\pm}=[2(1\pm e^{-2\alpha^2})]^{-1/2}$ is the
  normalization factor of the even/odd cat state [Eq.~(\ref{a4}) of
  the main text] at $r = 1$.}


{The product form (\ref{z3}) and the additive decomposition (\ref{z6})
  allow one to evaluate the $k$-dimensional integrals (\ref{z8}) by
  integrating sequentially from $y_k$ inward to $y_1$. \textcolor{black}{At the innermost step,}
  one encounters a single-variable integral of the form
\begin{gather}
\mathcal{E}(\eta,\mu,\delta;d) \equiv \frac{\eta}{\sqrt\pi}\int_{-d/2}^{d/2}e^{-\eta^2(y-\mu)^2+i\delta(y-\mu)}\,dy, 
\label{z12}
\end{gather}
which represents the postselection of one homodyne outcome within the
window $[-d/2, d/2]$ for a conditional Gaussian with inverse-width
parameter $\eta$, mean $\mu$, and interference frequency $\delta$.
Completing the square
in the exponent,
		\textcolor{black}{$-\eta^2(y-\mu)^2+i\delta(y-\mu)
		=-[\eta(y-\mu)-i\delta/(2\eta)]^2-\delta^2/(4\eta^2)$},
and substituting $t=\eta(y-\mu)-i\delta/(2\eta)$ with limits $t_\pm=
\eta(\pm d/2-\mu)-i\delta/(2\eta)$, we obtain
\begin{align}
  &
\label{z13}
\mathcal{E}(\eta,\mu,\delta;d) = \frac{e^{-\delta^2/(4\eta^2)}}{2}\times
  \notag
  \\
  &
\left[\mathrm{erf}\!\left(\eta\!\left(\tfrac{d}{2}{-}\mu\right)-\tfrac{i\delta}{2\eta}\right)+\mathrm{erf}\!\left(\eta\!\left(\tfrac{d}{2}{+}\mu\right)+\tfrac{i\delta}{2\eta}\right)\right], 
\end{align}
where
$\mathrm{erf}(z)$ denotes the error function of complex argument. The
prefactor
$e^{-\delta^2/(4\eta^2)}$ quantifies the suppression of the
interference contribution due to oscillatory cancellation within the
acceptance window. For $\delta\ne 0$ and $\mu\ne
0$, the function
$\mathcal{E}$ is in general complex-valued, which is the origin of the
$\mathrm{Re}$ operation in Eq.~(\ref{z7}). In the absence of
interference ($\delta=
0$), Eq.~(\ref{z13}) reduces to the real-valued expression
\begin{align}
  &
\label{z14}
\mathcal{G}(\eta,\mu;d)\equiv\mathcal{E}(\eta,\mu,0;d) = 
   \notag
  \\
  &
 \frac{1}{2}\left[\mathrm{erf}\!\left(\eta\!\left(\tfrac{d}{2}{+}\mu\right)\right)+\mathrm{erf}\!\left(\eta\!\left(\tfrac{d}{2}{-}\mu\right)\right)\right],
\end{align}
which is simply the probability that a Gaussian variable with inverse-width parameter $\eta$ and mean $\mu$ falls within $[-d/2, d/2]$.}

{The sequential evaluation proceeds as follows. \textcolor{black}{For
  $J_k^{(0)}$, the innermost integral over
  $y_k$ yields $\mathcal{G}(\eta_k,\mu_k;d)$, where $\mu_k$ depends only
  on the previously retained outcomes $y_1,\ldots,y_{k-1}$. This result
  is then multiplied by the $j=k-1$ Gaussian factor and integrated over
  $y_{k-1}$, and the procedure is repeated recursively:}
\begin{widetext}
\begin{gather}
J_k^{(0)}(d) = \int_{-d/2}^{d/2}\!\frac{\eta_1}{\sqrt\pi}e^{-\eta_1^2 y_1^2}\left[\int_{-d/2}^{d/2}\!\frac{\eta_2}{\sqrt\pi}e^{-\eta_2^2(y_2-\mu_2)^2}\left[\cdots\,\mathcal{G}(\eta_k,\mu_k;d)\cdots\right]dy_2\right]dy_1.
\label{z15}
\end{gather}
\end{widetext}
For
$J_k^{(1)}$, the procedure is analogous, with each Gaussian factor
acquiring the oscillatory phase
$e^{i\delta_j(y_j-\mu_j)}$ and the innermost building block replaced
by $\mathcal{E}$:
\begin{widetext}
\begin{gather}
J_k^{(1)}(d) = \int_{-d/2}^{d/2}\!\frac{\eta_1}{\sqrt\pi}e^{-\eta_1^2 y_1^2+i\delta_1 y_1}\left[\int\!\cdots\,\mathcal{E}(\eta_k,\mu_k,\delta_k;d)\cdots\,dy_2\right]dy_1. 
\label{z16}
\end{gather}
\end{widetext}
\textcolor{black}{Equations~(\ref{z15}) and (\ref{z16}) should therefore be understood as
nested conditional quadratures, each over a single homodyne outcome.}}

{Combining all results, we obtain the total probability for arbitrary
  $k$:
\begin{gather}
P_{(k)}(d) = \frac{J_k^{(0)}(d)+s\,e^{-a^2 G^{2k}/A_k}\;\mathrm{Re}\,J_k^{(1)}(d)}{1+s\,e^{-a^2}},
\label{z17}
\end{gather}
where $J_k^{(0)}$ and $J_k^{(1)}$ are given by the sequential
integrals (\ref{z15}, \ref{z16}), built from the elementary block
$\mathcal{E}$ (Eq.~\ref{z13}), with the protocol-defined parameters
$\eta_j$, $\mu_j$, $\delta_j$ (Eqs.~(\ref{z4}-\ref{z6})).}

{The evaluation proceeds as follows: (i) compute the parameters
  $r_{(j)}^2=\sum_{l=0}^j G^{2l}$, $\eta_j=r_{(j-1)}/r_{(j)}$, and
  $\delta_j=2a(-G)^{j-1}/r_{(j)}^2$ for $j=1,\ldots,k$; (ii) build the
  conditional means $\mu_j$ via the recursion $\mu_1=0$,
  $\mu_{j+1}=Gr_{(j-1)}^2(y_j-\mu_j)/r_{(j)}^2$; (iii) starting from
  $y_k$, evaluate the nested conditional quadratures inward, using
  $\mathcal{E}$ (or $\mathcal{G}$) for the innermost building block;
  and (iv) assemble the result via
  Eq.~(\ref{z17}) with $\mathcal{N}_k = 1+s\,e^{-a^2}$.}

{Next, we obtain an explicit result for $k=1$. For a single iteration the parameters are $\eta_1=1/r_{(1)}$, $\mu_1=0$, $\delta_1=2a/r_{(1)}^2$, $\mathcal{V}_1=e^{-a^2 G^2/r_{(1)}^2}$, and $r_{(1)}=\sqrt{1+G^2}$. No outer integration remains, and the building blocks give $J_1^{(0)}= \mathrm{erf}(d/(2r_{(1)}))$ and $\mathcal{V}_1J_1^{(1)}= e^{-a^2}\,\mathrm{Re}[\mathrm{erf}((d+2ia)/(2r_{(1)}))]$, leading to
\begin{widetext}
\begin{gather}
P_{(1)}(d) = \frac{1}{1+se^{-a^2}}\left[\mathrm{erf}\!\left(\frac{d}{2r_{(1)}}\right)+s\,e^{-a^2}\;\mathrm{Re}\left(\mathrm{erf}\!\left(\frac{d+2ia}{2r_{(1)}}\right)\right)\right].
\label{z18}
\end{gather}
\end{widetext}
In the limit $d\to\infty$ this gives $P_{(1)}(\infty)=1$. For the even
cat, the limit $a\to0$ (vacuum ancilla) reduces the expression to
$P_{(1)}(d)=\mathrm{erf}(d/(2r_{(1)}))$. The normalized odd-parity
state is singular at $a=0$ and should therefore be treated separately
in that limit. For large $a$ the interference term is exponentially
suppressed and the success probability is dominated by the Gaussian
contribution.}

\end{document}